\newcommand\oiii{[\ion{O}{3}]$\lambda$5007\AA}
\newcommand\ujy{$\mu$Jy}

\documentclass[iop,revtex4]{emulateapj}
\usepackage{color,lscape}

\shorttitle{Sensitive radio survey of obscured quasar candidates}
\shortauthors{Alexandroff et al.}

\begin{document}
\title{Sensitive radio survey of obscured quasar candidates} 

\author{Rachael M. Alexandroff\altaffilmark{1}, Nadia L. Zakamska\altaffilmark{2,1}, Sjoert van Velzen\altaffilmark{1}, Jenny E. Greene\altaffilmark{3}, Michael A. Strauss\altaffilmark{3}}
\altaffiltext{1}{Center for Astrophysical Sciences, Department of Physics and Astronomy, Johns Hopkins University, Baltimore, MD 21218, USA}
\altaffiltext{2}{D. Lunder and A. Ezekowitz Junior Visiting Professor, Institute for Advanced Study, Einstein Dr., Princeton, NJ 08540, USA} 
\altaffiltext{3}{Department of Astrophysical Sciences, Princeton University, Princeton, NJ 08544, USA}
\email{rmalexan@jhu.edu}

\label{firstpage}

\begin{abstract}
We study the radio properties of moderately obscured quasars over a range of redshifts to understand the role of radio activity in accretion using the Jansky Very Large Array (JVLA) at 6.0GHz and 1.4GHz. Our $z\sim 2.5$ sample consists of optically-selected obscured quasar candidates, all of which are radio-quiet, with typical radio luminosities of $\nu L_{\nu}$[1.4 GHz]$\la 10^{40}$ erg s$^{-1}$. Only a single source is individually detected in our deep (rms$\sim 10$ \ujy) exposures. This population would not be identified by radio-based selection methods used for distinguishing dusty star-forming galaxies and obscured active nuclei.  In our pilot A-array study of $z \sim 0.5$ radio-quiet quasars, we spatially resolve four of five objects on scales $\sim 5$ kpc and find they have steep spectral indices. Therefore, radio emission in these sources could be due to jet-driven or radiatively driven bubbles interacting with interstellar material on the scale of the host galaxy. Finally, we also study the population of $\sim 200$ faint ($\sim 40\mu$Jy - 40mJy) radio sources observed over $\sim 120$ arcmin$^2$ of our data.  60\% of these detections are matched in the SDSS and/or WISE and are, in roughly equal shares, active nuclei at a broad range of redshifts, passive galaxies with no other signs of nuclear activity and IR-bright but optically faint sources. Spectroscopically or photometrically confirmed star-forming galaxies constitute only a small minority of the matches. Such sensitive radio surveys allow us to address important questions of AGN evolution and evaluate the AGN contribution to the radio-quiet sky.
\end{abstract}

\section{Introduction}
\label{sec:intro}

The composition of the sub-mJy radio sky, including the active galactic nuclei (AGN) fraction at sub-mJy levels, is still an open question as it is only recently that the capabilities of the expanded JVLA, $5-20 \times$ more sensitive than the original VLA, have opened up the realm of sub-mJy radio populations without stacking analysis.  A change in the overall slope of differential radio source counts below $\sim1$ mJy suggests that a new population of sources begins to contribute below these flux densities.  Theoretical predictions \citep[e.g][]{Massardi2010} and observations by e.g. \citet{Bonzini2013} suggest that at the level of hundreds of \ujy\ the radio sky begins to be dominated by star-forming galaxies. Alternatively, \citet{Jarvis2004} have also suggested, based on X-ray source counts, that type 2 and low luminosity AGN could contribute most of the necessary radio flux below 1 mJy.  

Recent deep surveys such as VLA-COSMOS \citep{Schinnerer2004} and the Extended Chandra Deep Field South \citep[E-CDFS][]{Miller2013} have differed slightly in their detection fraction of multi-wavelength counterparts (essential for proper source classification), their classification schemes and results.  In the E-CDFS, which covered $\approx 0.3$ deg$^2$ down to an average 5$\sigma$ flux density of $\approx 37$ \ujy\ with the Jansky Very Large Array (JVLA) at 1.4 GHz, \citet{Bonzini2013} found that AGN made up $43\%$ of their entire sample of 883 sources, being $100\%$ of sources at $\sim 10$ mJy but only $38\%$ at the survey limit with the remainder of sources being star-forming (SF) galaxies.  Meanwhile, \citet{Smolcic2008} find a nearly constant combined fraction of 70\% for AGN and quasars in the VLA-COSMOS Survey, covering 2 deg$^2$ in the COSMOS field down to the survey flux limit of 50 \ujy.

At a given bolometric AGN luminosity, the observed radio power varies over many orders of magnitude, with only a small fraction ($\approx 15-10\%$) of the most luminous optically-selected sources displaying classical radio-loud jets on kiloparsec scales \citep{Kellermann1989,Xu1999,Zakamska2004}. Classically, radio sources are divided based upon the ratio of their radio luminosity to optical luminosity (or ``radio-loudness"), though there is continued disagreement as to whether these two populations, ``radio-loud" and ``radio-quiet" sources represent a true dichotomy \citep[e.g.][]{Kellermann1989,Ivezic2002,Dunlop2003,White2007,Bonchi2013}. If there is a break or bimodality in the radio luminosity function of quasars \citep{Kimball2011,Condon2013}, this might suggest that two different mechanisms may be responsible for the radio emission in the radio-loud and radio-quiet quasar populations. 

It is not clear what mechanism would produce the radio emission in radio-quiet quasars. Currently there are four working hypotheses: (i) radio-quiet quasars are simply the scaled-down version of their radio-loud counterparts, implying that the radio emission observed is from a compact jet \citep{Kukula1998,Ulvestad2005,Giroletti2009}. (ii) Radio emission could also be due to a synchrotron emission in accretion disk coronae\citep{Laor2008}. (iii) The presence of a relationship between the kinematics of ionized gas in local Type 2 radio-quiet quasars and their radio luminosity \citep{Mullaney2013,Zakamska2014} suggests that quasar-driven outflows could be the source of the radio emission in radio-quiet quasars \citep{Stocke1992,FG2012,Zubovas2012,Zakamska2014,Nims2015}.(iv) Finally, radio emission associated with star formation in the host galaxy may be responsible for most or all of the faint radio luminosity in these sources \citep[e.g.][]{Kimball2011,Padovani2011,Condon2013}.

Differentiating between these scenarios is difficult. Some arguments can be made on the grounds of energetics; for example, star formation rates in host galaxies of quasars at $z<1$ are inadequate to explain the observed radio emission by about an order of magnitude \citep{Zakamska2016a}, though the same argument might not apply to lower luminosity AGN \citep{Rosario2013} or quasars at $z>2$ \citep{Kimball2011,Condon2013}. As for scenarios (i)$-$(iii), a detailed analysis of radio spectral indices ($F_{\nu} \propto \nu^{\alpha}$) and radio morphology for a large sample of quasars presents the best path for differentiating between these possible explanations for the radio emission in radio-quiet quasars. In particular, coronal emission is expected to be on parsec scales and have flat spectral indices ($\alpha \approx 0$; \citealt{Laor2008}), similar to parsec-scale cores of radio jets, whereas radio emission produced on larger scales either in winds or in jet-powered lobes or in star formation is expected to have steep spectra ($\alpha \approx -0.7$) and be partially resolved. For the first time, the resolution and sensitivity capabilities of the JVLA makes such a dedicated study possible. 

Here we present the results of two separate initial surveys of the sub-mJy radio sky taken with the JVLA in A and B-configurations.  The data for this project were intended to study the continuum radio properties of radio-radio-intermediateintermediate and radio-quiet predominantly optically-obscured quasars at both low ($z \lesssim 0.8$) and high ($z \sim 2.5$) redshifts using observations that were at least $5\times$ deeper than the Faint Images of the Radio Sky at Twenty Centimeters \citep[FIRST;][]{Becker1995} survey. In \S~\ref{sec:sample} we describe our various sample selection methods as well as the JVLA observations and subsequent data reduction and analysis. We look more closely at our samples of high and low-redshift radio-quiet obscured quasars in \S~\ref{sec:highz} and \S~\ref{sec:lowz}. Then \S~\ref{sec:sourcepop} explores the sample properties of the sub-mJy radio sources we have identified as well as their optical and mid-infrared (MIR) counterparts and we present some interesting sources in \S~\ref{sec:interesting}. Finally, we discuss the implications of our results and offer our conclusions in \S~\ref{sec:conclusion}.  

We use a $h=0.7, \Omega_m=0.3, \Omega_{\Lambda}=0.7$ cosmology throughout the paper. We quote radio luminosities k-corrected to the rest frame 1.4 GHz for ease of comparison with other datasets using equation
\begin{equation}
\nu L_{\nu}{\rm [1.4 GHz]}=4 \pi D_L^2 (1+z)^{-1-\alpha}\left(\frac{\nu}{\nu_{\rm obs}}\right)^{1+\alpha}\nu_{\rm obs} F_{\nu,{\rm obs}}.
\label{eq_lum}
\end{equation}
Here, $\nu$ and $L_{\nu}$ are at 1.4 GHz, and $\nu_{\rm obs}$ and $F_{\nu,{\rm obs}}$ are at the frequency of our observations (1.4 or 6 GHz), and for k-corrections we use spectral index $\alpha=-0.7$ when it was not possible to calculate the observed value. We observe objects at $z\sim 2.5$ at 1.4 and 6 GHz, corresponding to rest-frame frequencies 5.3 and 21 GHz, and therefore the mismatch between our reference frequency (1.4 GHz) and the rest-frame frequency probed by our observations is quite significant. If the spectral index is $\alpha=-0.3$ instead of $-0.7$ for a $z=2.5$ source whose flux is measured at $\nu_{\rm obs}=6$ GHz, its true 1.4 GHz intrinsic luminosity is 3 times fainter than that inferred from equation (\ref{eq_lum}). In \S~\ref{sec:sample}$-$\S~\ref{sec:lowz}, we use SDSS~Jhhmm+ddmm notation for identifying our primary targets (the centers of our fields) and in \S~\ref{sec:sourcepop}-\S~\ref{sec:interesting} we use the full SDSS~Jhhmmss.ss+ddmmss.s notation for identifying sources in the field. 

\section{Sample Selection, Observations and Data Reduction}
\label{sec:sample}

In this section we describe the sample selection, the observations, steps of the data reduction and analysis performed on each field.

\subsection{Sample Selection and Observations}

Our high-redshift program VLA/13B-382 targeted a sample of 11 high-redshift ($2.0 < z < 4.2$) obscured quasar candidates from \citet{Alexandroff2013}.  These sources were originally selected from the Sloan Digital Sky Survey (SDSS) Baryon Oscillation Spectroscopic Survey \citep[BOSS;][]{Dawson2013} by their narrow emission line widths (FWHM $< 2000$ km s$^{-1}$ in both \ion{C}{4} and Ly$\alpha$) and weak continuum in the rest-frame UV. Only objects from Data Release 9 \citep{Ahn2012} or earlier are included in this search due to the timeline of the research.  Follow-up observations of a sub-sample of twenty-five of these objects in the rest-frame optical \citep{Greene2014} showed that most of them have a broad H$\alpha$ component and that therefore most of these objects must have intermediate values of extinction ($0 < A_{\rm V} < 2.2$ mag) akin to type 1.8/1.9 quasars.  The basic radio properties of this population were presented by \citet{Alexandroff2013}. 

We chose eleven objects for deep observations with the JVLA (see Table \ref{tab:highz}), selected to have the greatest possible overlap with other multi-wavelength follow-up of the original \citet{Alexandroff2013} sample; none were detected in FIRST. This sample was observed with the JVLA in both L- and C-bands ($\approx$ 1$-$2 GHz and $\approx$ 4$-$8 GHz respectively) in the B configuration with $\sim 4.3$\arcsec\ and $\sim 1$\arcsec\ resolution respectively. The spatial resolution at $\sim 6$ GHz corresponds to a physical scale of $\sim 8$ kpc at $z \sim 2.5$.

Observations were scheduled dynamically in blocks of three objects.  We observed one flux/bandpass calibrator at the beginning of each observation set.  Targets were observed while nodding between a phase/amplitude calibrator ever twenty minutes.  Total on-source time was 32.5 minutes in the L-band and 28 minutes in the C-band per object. In the L-band, we recorded in full polarization the total 16 contiguous spectral windows with $64 \times 1$ MHz channels each to yield a total instantaneous bandwidth of 1024 MHz centered at 1.4 GHz.  In the C-band we had two frequency bands centered at $\approx 5$ GHz and $\approx 7$ GHz respectively, each with 8 spectral windows of $64 \times 2$ MHz channels to yield a total of 2048 MHz bandwidth.

Our low-redshift program VLA/14A-310 originally targeted 106 type 2 and type 1 radio-quiet quasars at $z\lesssim 0.8$ from \citet{Reyes2008} and \citet{Liu2014}, but was scheduled as filler, and as a result only five sources were observed (Table \ref{tab:lowz}): four type 2 quasars and one type 1. Of the original sample of 106 objects proposed, 80$\%$ were detected with FIRST at no more than a few mJy level. Of the five objects observed, all but one were previously detected in FIRST and only one of the sources in FIRST (SDSS1123+3105) was clearly resolved. 

This sample was observed in the C-band ($\approx$ 4$-$8 GHz) in the A configuration which provides resolution of 0.33\arcsec, corresponding to $\sim 2$ kpc at $z\sim 0.5$. All objects were observed on April 14, 2014. Based on our experience with the high-redshift program, we adjusted central frequencies of the two bands to 5.25 and 7.2 GHz, which reduced somewhat the effects of radio frequency interference. Observations included the flux standard 3C286 and were conducted with nodding to an appropriate amplitude and phase calibrator.  Total on-source time was 11.5 minutes per source.

\subsection{Data Reduction and Analysis}

We reduced the data using the Common Astronomy Software Applications (CASA) package v4.3.0 \citep{McMullin2007}. Raw visibilities were calibrated using the VLA Calibration pipeline version 1.3.1\footnote{https://science.nrao.edu/facilities/vla/data-processing}. All solutions were inspected and additional flagging, as necessary, was accomplished by hand using CASA's {\sl Plotms} task.

All maps were created in CASA at the band center with a Briggs weighting scheme of ROBUST = 2.0 (natural weighting) to maximize our sensitivity to faint sources.  Map size was set to match the primary beam full-width at half power 
which is approximately 7 arcminutes in the C-band.  Certain fields included strong sources far from the image center that left residuals in the images which were difficult to clean.  In these cases larger maps were created for the purpose of cleaning but the final images analyzed were cropped to be the same size as the rest of the sample. Finally, we corrected every field for primary beam attenuation using the task {\sl pbcorr}.  The typical rms at the field center was $\sim 1.5 \times 10^{-5}$Jy/beam for the C-band and $\sim8 \times 10^{-5}$Jy/beam in the L-band.


\subsection{CLEAN bias}
\label{ssec:CLEAN}

CLEAN bias is a loss of flux due to sparse uv-coverage \citep{White1997, Condon1998}. To measure the CLEAN bias we insert two fake point sources (of flux 1 mJy and 0.1 mJy) into two of the 6.0GHz B-array fields (SDSS~J2233+0249 and SDSS~J0046+0005) at the field center. We then clean on the fake source position and measure the flux. We find an average CLEAN bias of $-12$ \ujy\ at 6 GHz. At 1.4GHz we insert a fake point source of flux 10mJy at the field center of SDSS~J0046+0005 and measure a CLEAN bias of 1.1mJy. Therefore, CLEAN bias is not important for those of our targets whose fluxes are known from FIRST to be $\ga 1$ mJy. We take the CLEAN bias into account when evaluating the quality of spectral indices in the faint radio population (\S~\ref{sec:sourcepop}).

\section{High redshift quasars}
\label{sec:highz}

In this section, we discuss the JVLA observations of eleven moderately obscured quasar candidates at $z=2-3$ (Table \ref{tab:highz}) and the implications of these data in \S~\ref{ssec:h_imp}. These objects are selected based on their emission line properties from the SDSS spectroscopic database \citep{Alexandroff2013} and then shown to be moderately obscured ($A_{\rm V}\la$ a few mag) using follow-up near-infrared spectroscopy \citep{Greene2014}. 

\begin{deluxetable}{l|r|r|r|r}
\tablecolumns{11}
\tablecaption{High Redshift Sample Properties\label{tab:highz}}
\tablehead{
Source name, & $z$ & $F_{\nu}^{\rm peak}$, & $F_{\nu}^{\rm rms}$, & FWHM \\
SDSS coordinates &  & 6 GHz & 6 GHz & [OIII] }
\startdata
SDSS~J004600.48+000543.65 & 2.458 & $<7.07$ & 9.36 & \\
SDSS~J004728.77+004020.30 & 3.063 & $<5.41$ & 9.47 & \\
SDSS~J013327.23+001959.61 & 2.723 & $<19.2$ & 5.08 & \\
SDSS~J090612.64+030900.37 & 2.503 & $<19.9$ & 11.9 & \\
SDSS~J091357.87+005530.72 & 3.206 & $<25.8$ & 10.4 & \\
SDSS~J091301.33+034207.60 & 3.006 & $<29.3$ & 10.5 & $390\pm40$\\
SDSS~J095118.93+450432.42 & 2.451 & $<23.8$ & 10.2 & $401\pm50$\\
SDSS~J103249.55+373649.03 & 2.354 & $<8.85$ & 13.1& $840\pm410$\\
SDSS~J220126.09+001231.50 & 2.635 & 268.0 & 7.63& \\
SDSS~J222946.61+005540.51 & 2.368 & $<8.51$ & 4.69 & $550\pm50$\\
SDSS~J223348.07+024932.80 & 2.587 & $<3.73$ & 5.20 & \\
\enddata
\tablecomments{Peak fluxes and rms values are in $\mu$Jy/beam and are measured using {\sl Aegean} to do forced measurements at the optical locations of our sources.  SDSS~J2201+0012 is the only real detection. FWHM[OIII] is the full width at half maximum of the [OIII]$\lambda$5007\AA\ emission line where available from the near-infrared follow-up observations, in km s$^{-1}$ \citep{Greene2014}.}
\end{deluxetable}

\subsection{FIRST and JVLA observations of high-redshift obscured quasar candidates}
\label{ssec:h_obs}

In \citet{Alexandroff2013} we matched our sample of 145 high redshift Type 2 quasar candidates with the FIRST all-sky survey.  We found a detection rate of only 2\% (3 matches with FIRST within 3\arcsec), low by comparison to the radio-loud fraction of quasars, which (depending on the definition of radio-loudness) is estimated at $\sim 10-20\%$ \citep{Zakamska2004,Jiang2007,Kratzer2015}. Of the three sources detected in FIRST, all have measured peak fluxes of only $\sim 2$ mJy (accounting for CLEAN bias) which corresponds to a  \mbox{k-corrected} luminosity of $\nu L_{\nu}$[1.4 GHz] $= 8.5 \times 10^{41}$ erg s$^{-1}$ at a mean redshift of $\langle z \rangle = 2.34$. At this redshift, such objects may be considered radio-intermediate \citep{Xu1999}.  But with typical luminosity sensitivity at this redshift of 10$^{42}$ erg s$^{-1}$, the FIRST survey is not quite deep enough to probe the transition between the radio-quiet and radio-loud population, and deeper observations were required

\begin{figure*}
\includegraphics[scale=0.5,trim=120 350 0 100,clip=true]{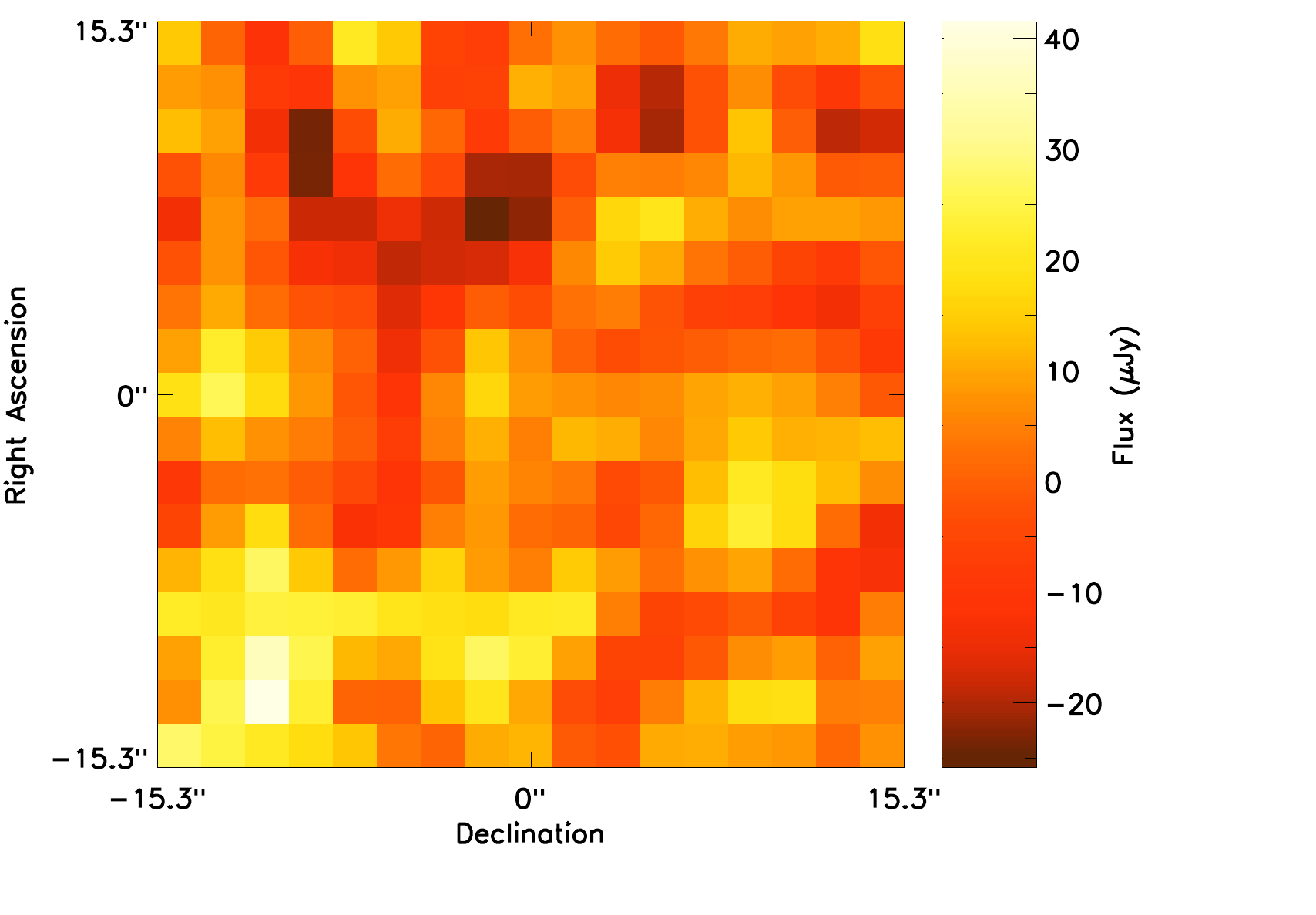}
\includegraphics[scale=0.5,trim=120 350 0 100,clip=true]{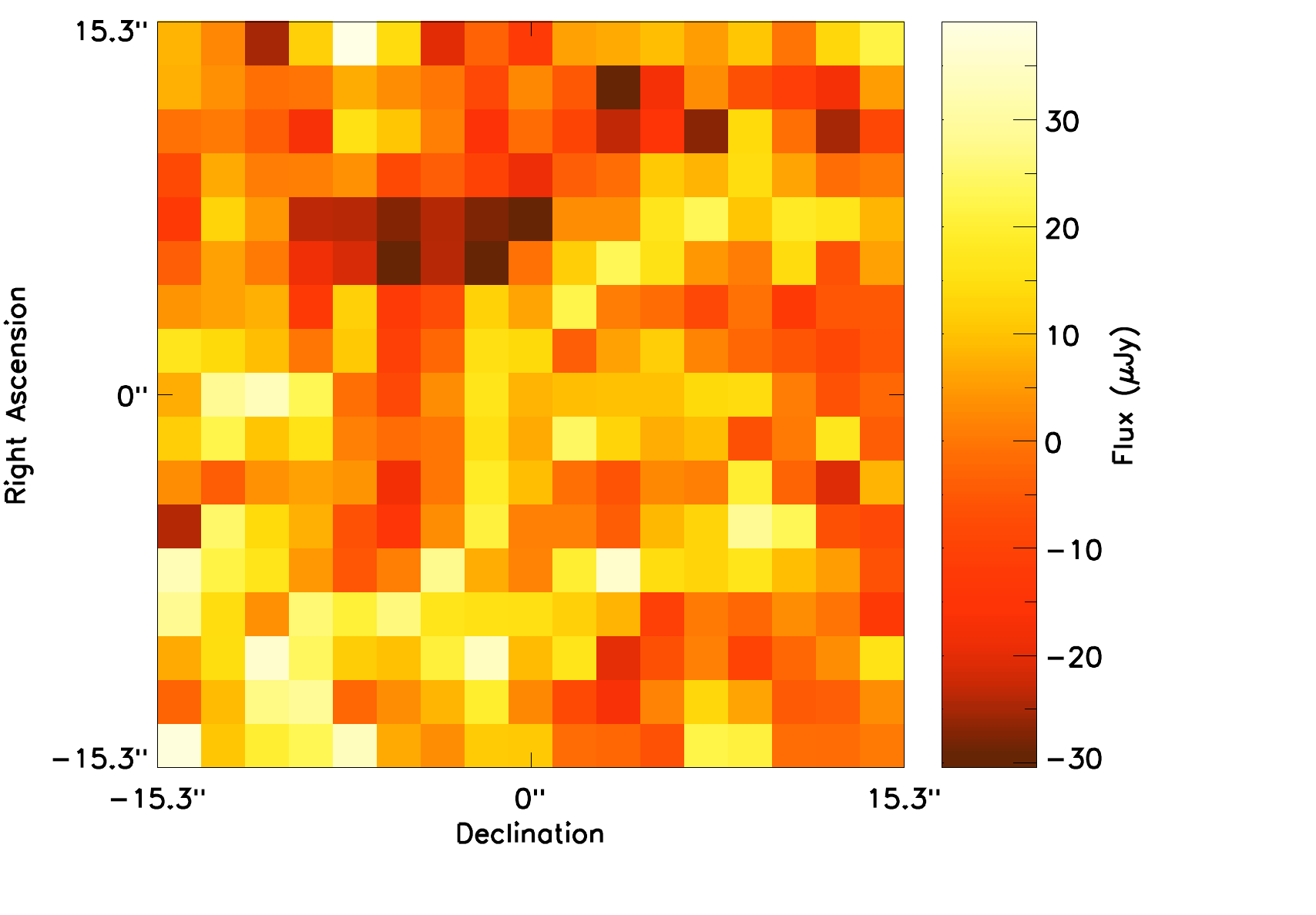}
\caption{Mean (left) and median (right) stacks of FIRST images (20 cm) at the locations of all 142 non-detected Type 2 quasar candidates from \citet{Alexandroff2013} displayed using asinh scaling. Nothing is detected in either stack, allowing us to set an upper limit on the mean flux of our Type 2 sources at 35 \ujy (without Snapshot bias).}
\label{pic:stack_FIRST}
\end{figure*}

We start our analysis of the radio properties of this population with a stack of all 142 non-detected sources from the FIRST survey. We extract $30.6\arcsec \times 30.6\arcsec$ cutouts from the FIRST survey\footnote{http://third.ucllnl.org/cgi-bin/firstcutout} at the SDSS position of each quasar that was not detected at the FIRST survey limit of $\sim1$ mJy and combine them using both a mean and median stacking (see Figure \ref{pic:stack_FIRST}). We find no detection in either stack. The standard deviation of the mean image is 11.7 \ujy\ and so we place the upper limit for a detection at 3$\sigma$ or 35 \ujy. While these sources were below the FIRST detection threshold, and thus were not CLEANed during FIRST image processing, we must still account for Snapshot bias in our detection which is also believed to be the result of the non-linear CLEAN processing of images though it is still not well understood \citep{White2007}.  We correct for Snapshot bias using equation (1) in \citet{White2007} obtaining a 49 \ujy\ upper limit on the mean flux of our stack. At a redshift of 2.5 our measured upper limit on the flux corresponds to a k-corrected luminosity of $\nu L_{\nu}$[1.4 GHz] $< 2.4 \times 10^{40}$ erg s$^{-1}$. 

\begin{figure}
\includegraphics[scale=0.5,trim=120 350 0 100,clip=true]{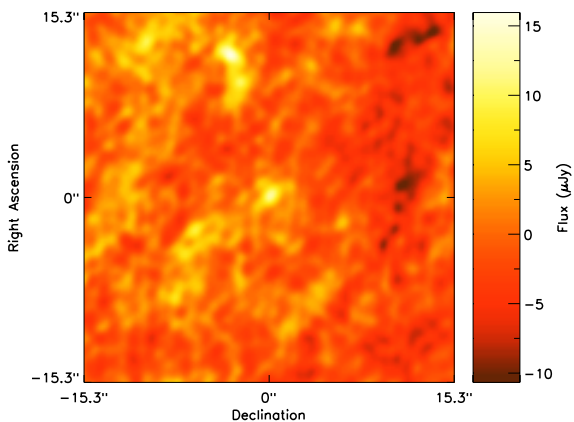}
\caption{Mean stack of JVLA images at the locations of all non-detected Type 2 quasar candidates from our JVLA program. We see a clear detection here with a mean flux of 12.5 \ujy\ (not accounting for CLEAN bias). This detection is also seen in additional stacks in which we remove a single source at a time indicating that the radio power is not dominated by a single bright source. The other detection, on the top left of the image, comes from a bright source in the field of SDSS~J0818+2237.}
\label{pic:stack_VLA}
\end{figure}

In our JVLA program we targeted eleven high redshift obscured radio-quiet quasar candidates from \citet{Alexandroff2013}. Only one, SDSS~J2201+0012, is detected in our JVLA program with a flux of 0.244 mJy corresponding to a radio power of $\nu L_{\nu}$[1.4 GHz] $= 8.5 \times 10^{40}$ erg s$^{-1}$ at $z=2.6$. This is above the limit set by our FIRST stacking analysis. We extract cutouts of the same size as the FIRST cutouts from our JVLA maps around the SDSS positions of our undetected sources at 6.0 GHz. We see a detection in the mean stack with a flux of 12.5 \ujy\ at 6 GHz (see Figure \ref{pic:stack_VLA}) which, correcting for our measured CLEAN bias, corresponds to an expected real flux of $\sim 24.5$ \ujy. This corresponds to a radio power (using the mean redshift of our sample $\langle z \rangle = 2.67$) of $\nu L_{\nu}$[1.4 GHz]$=9.1 (\pm1.1) \times 10^{39}$ erg s$^{-1}$. We make additional stacks, removing one source in each stack, and find that each shows a clear detection, so the radio power in our stack is not predominantly from a single object. The flux of the stacked detection, if we assume a spectral index of $\alpha=-0.7$, is just below the upper limit set by stacking our FIRST images.  

\subsection{Implications}
\label{ssec:h_imp}

Obscured quasar candidates at $z=2-3$ remain a challenging population to identify even in deep multi-wavelength surveys, and the full extent of this population remains unknown. Their high infrared-to-optical ratios can be mimicked by those of dusty star-forming galaxies \citep{Martinez2006a}. Optical colors alone do not distinguish these objects from various other populations \citep{Alexandroff2013} because strong narrow emission lines with varying equivalent widths combined with varying degree of extinction can yield a wide range of optical colors. Infrared color selection methods and infrared-to-optical color selection methods recover interesting populations of high-redshift obscured quasars \citep{Assef2015, Ross2015}, but it is not known whether these methods are complete even at the highest luminosities. 

To select a sample of high-redshift obscured quasar candidates, \citet{Martinez2006a} used radio fluxes as a distinguishing characteristic to separate quasar candidates from star-forming galaxies. Their minimal radio flux cut was 0.35 mJy at 1.4 GHz, corresponding to $\nu L_{\nu}$[1.4 GHz]$=1.7\times 10^{41}$ erg s$^{-1}$ at $z=2.5$. None of our sources would be uncovered with these observations, indicating that radio-based selection methods for obscured quasars recover only ``the tip of the iceberg" of the obscured quasar population.

As we've described above, the origin of the radio emission in radio-quiet obscured quasars is unknown, but it appears that the amount of this emission in AGN and quasars is positively correlated with the kinematics of the forbidden emission lines such as [OIII]$\lambda$5007\AA\ \citep{Mullaney2013, Zakamska2014}, implying that there is a connection between radio emission and the ionized gas winds driven by the quasar. The best-fit quadratic relationship is 
\begin{eqnarray}
\log(\nu L_{\nu}[1.4\mbox{ GHz}],\mbox{erg s}^{-1})=\nonumber\\
2\times\log(\mbox{FWHM, km s}^{-1})+34.47=\nonumber\\
2\times\log(\mbox{FWQM, km s}^{-1})+34.07,\label{eq:fwhm}
\end{eqnarray} 
with a standard deviation of around 0.5 dex around this relationship when the radio-loud quasars are removed. Here FWHM and FWQM are the full width at half maximum and at quarter maximum of the [OIII]$\lambda$5007\AA\ emission line. 

The strongest forbidden line [OIII]$\lambda$5007\AA\ is redshifted out of the optical spectrum at the redshifts of our targets and therefore is not directly accessible. To probe where our candidates lie in relation to this correlation, we use [OIII] kinematics analysis from \citet{Greene2014} who followed up 25 candidate obscured quasars with near-infrared observations, including four of the objects with JVLA observations (SDSS~J0913+0342, SDSS~J0951+4504, SDSS~J1032+3736, and SDSS~J2229+0055). The median FWHM([OIII]) in that sample is 475 km s$^{-1}$, which would imply the median radio luminosity of $\nu L_{\nu}$[1.4 GHz]$=10^{39.8}$ erg s$^{-1}$, entirely consistent with the observed value ($10^{40.0}$ erg s$^{-1}$). The objects in our sample are selected in a manner similar to the optical selection of low redshift ($z\sim0.5$) obscured quasars \citep{Zakamska2003,Reyes2008}: on the basis of strong narrow rest-frame ultra-violet emission lines and the absence of a detectable quasar continuum.  When they are observed in the rest-frame optical they tend to show relatively quiescent [OIII] kinematics \citep{Greene2014} in addition to their small radio luminosities. 

We now compare these results with the properties seen in a population of extremely red quasars which are selected from BOSS and WISE based on their high infrared-to-optical ratios and high equivalent widths of the CIV$\lambda$1549\AA\ emission \citep{Ross2015}. Although these objects do not conform to all the classical characteristics of type 2 quasars, their multi-wavelength properties are indicative of large amounts of obscuration \citep[][Hamann et al. in prep.]{Ross2015}.  In follow-up near-infrared spectroscopy, four of these objects show [OIII]$\lambda$5007\AA\ with extremely high velocity widths, FWHM([OIII])=2800$-$5000 km s$^{-1}$ \citep{Zakamska2016b}. Radio stacking of a sample of 81 extremely red quasars using FIRST observations shows a mean flux of 0.13 mJy (Hamann et al. in prep.) which we estimate is about 0.18 mJy when corrected for the Snapshot bias \citep{White2007}.  Because this sample is of similar size to our type 2 sample which showed no detection, we conclude that there is a clear difference in the average radio properties of these two samples. 

This flux at the median redshift ($\langle z \rangle=2.48$) of the stacked sample of extremely red quasars corresponds to $\nu L_{\nu}$[1.4 GHz]$=8\times 10^{40}$ erg s$^{-1}$. With a median FWHM([OIII])$=2930$ km s$^{-1}$ \citep{Zakamska2016b}, we would expect a somewhat higher radio luminosity, $10^{41.4}$ erg s$^{-1}$. However, the objects picked for the near-infrared follow up thus far are the most extreme and therefore likely have a higher than average FWHM([OIII]) compared to the full sample, biasing any calculation of the expected radio luminosity from the radio/kinematics relationship.  Overall, the qualitative agreement between the low-redshift kinematics/radio relationship and the two high-redshift samples -- type 2 quasar candidates from this paper and extremely red quasars from Hamann et al. (in prep.) -- is remarkable and gives further support to the notion that radio emission in radio-quiet quasars is intimately connected with quasar-driven ionized gas outflows.

Radio emission in nearby radio-quiet AGN is often attributed to star formation\citep{Rosario2013}. However, in quasars with $L_{\rm bol}\ga 10^{45}$ erg s$^{-1}$ at low redshifts ($z\la 1$), radio emission due to star formation (as estimated from infrared measures of star formation) is about an order of magnitude too low to account for the observed radio luminosity \citep{Zakamska2016a}. In high-redshift quasars, hundreds of $M_{\odot}$ yr$^{-1}$ of star formation would be required to power the observed radio emission at the radio-quiet tail of the quasar distribution \citep{Kimball2011, Condon2013}. In the two classes of obscured quasars discussed here -- narrow-line selected type 2 quasar candidates and extremely red quasars -- 360 and 3800 $M_{\odot}$ yr$^{-1}$ respectively would be required to power all of the observed radio emission (our adopted calibration is discussed in \S~\ref{ssec:sfr}). We do not yet have any additional information to support or rule out this scenario of extremely high star formation accompanying the quasar activity in these objects. \citet{Tsai2015} argue against such high star formation rates for hot dust-obscured galaxies (which are analogs of extremely red quasars) on the basis of the lack of a molecular gas reservoir.  

\section{Low redshift quasars}
\label{sec:lowz}

As was disscused in the previous section, the radio emission of radio-quiet quasars is related to the ionized gas kinematics \citep{Mullaney2013, Zakamska2014}. With this in mind we discuss high-resolution A-array observations of five low-redshift quasars and their ionized gas kinematics in \S~\ref{ssec:lobs} and the implications of our observations in \S~\ref{ssec:limp}. 

\subsection{FIRST and JVLA observations of $z<1$ quasars}
\label{ssec:lobs}

\begin{figure}
\includegraphics[scale=0.55,trim=80 350 0 100,clip=true]{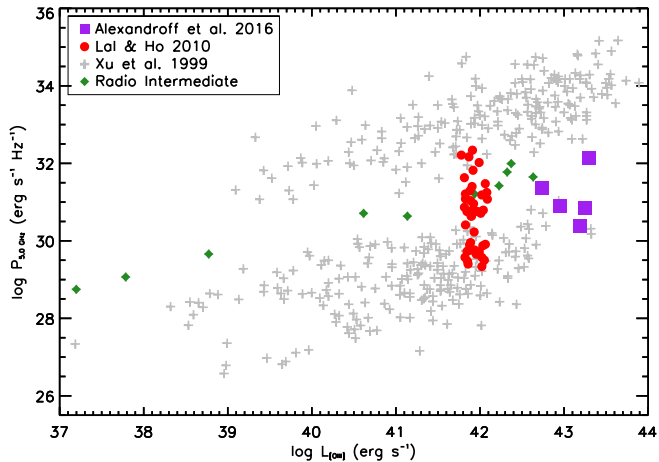}
\caption{Relationship between radio power and [OIII] luminosity for type 1 and type 2 AGN, after \citet{Xu1999} and \citet{Lal2010}. The black crosses are the sample of \citet{Xu1999} but using a modern cosmology. The green points are those of the objects from \citet{Xu1999} that are classified as radio-intermediate based on this diagram. The red circles are from the sample of \citet{Lal2010}. The five low-redshift Type 2 quasars observed in this program are plotted in purple.}
\label{pic:radio_power}
\end{figure}

We observed five fields with the A-array, four centered on type 2 quasars from \citet{Reyes2008} and one centered on a type 1 quasar from \citet{Liu2014}. All five are easily detected at 6~GHz in our A-array JVLA program. All of our objects lie in the radio-quiet or radio-intermediate regime based on the ratio of their \oiii\ emission to radio flux (Figure \ref{pic:radio_power}, after \citealt{Xu1999}). 

\begin{figure*}
\includegraphics[scale=0.18]{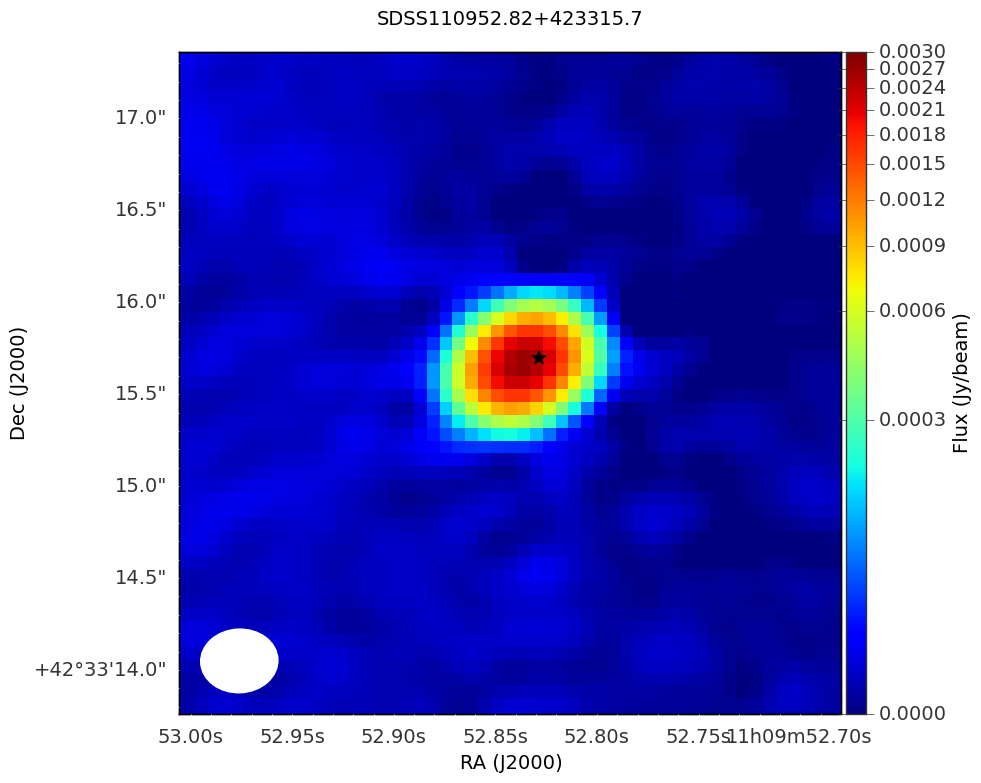}%
\includegraphics[scale=0.18]{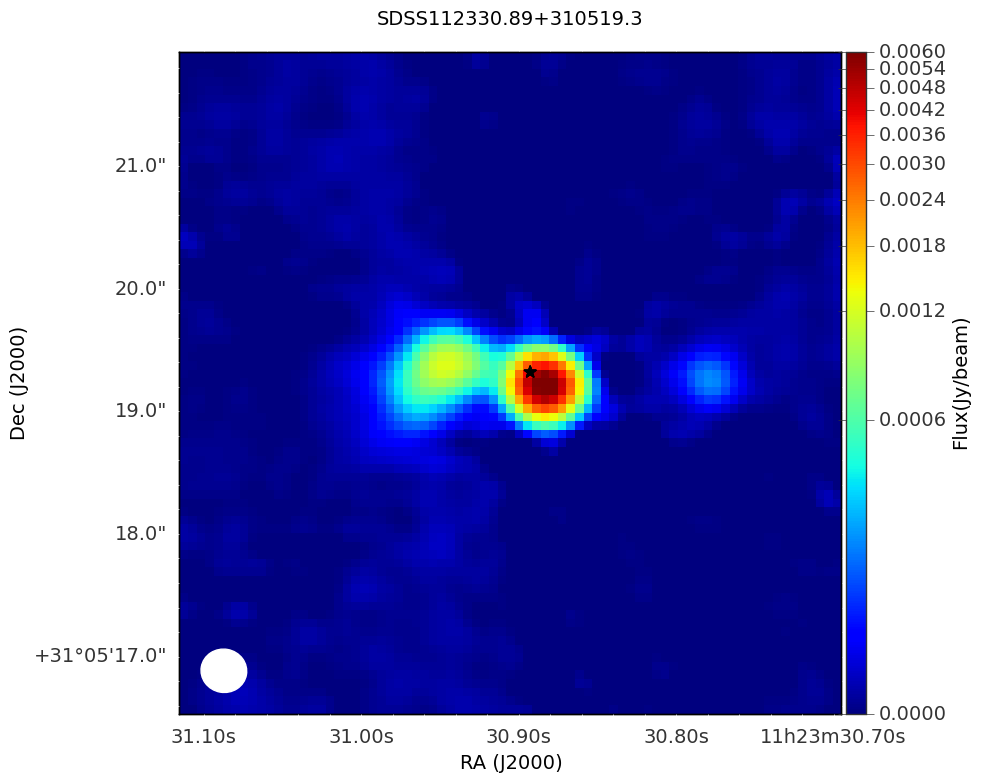}%
\includegraphics[scale=0.18]{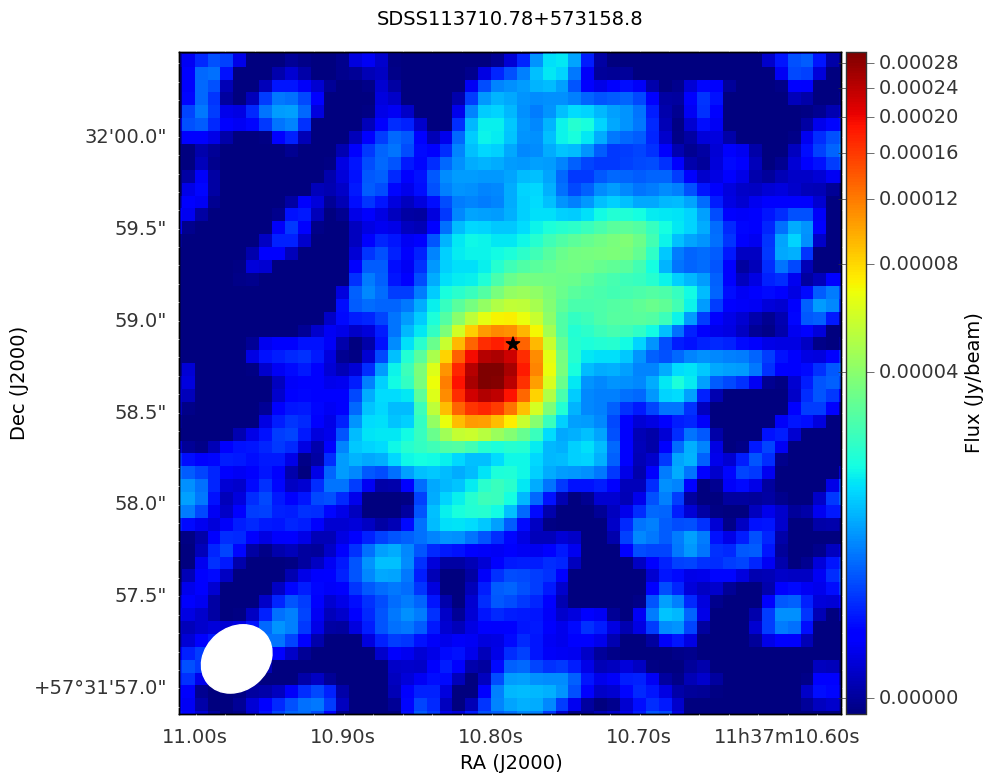}%
\includegraphics[scale=0.18]{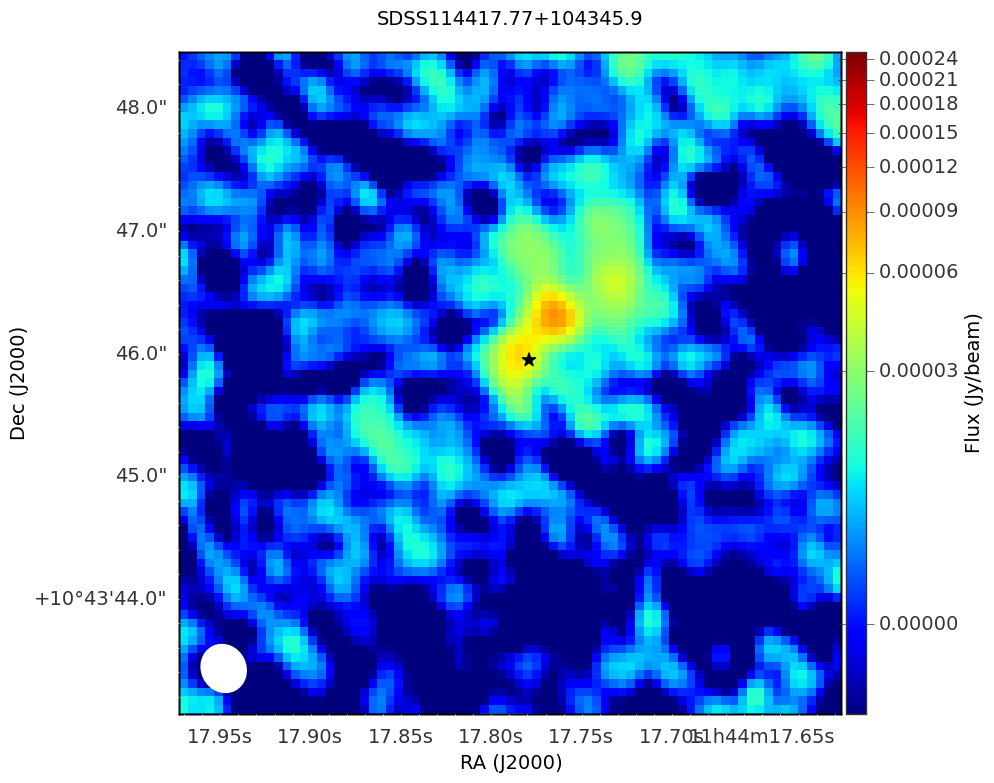}\\
\includegraphics[scale=0.44, clip=true, trim=0cm 0.6cm 10.5cm 10cm]{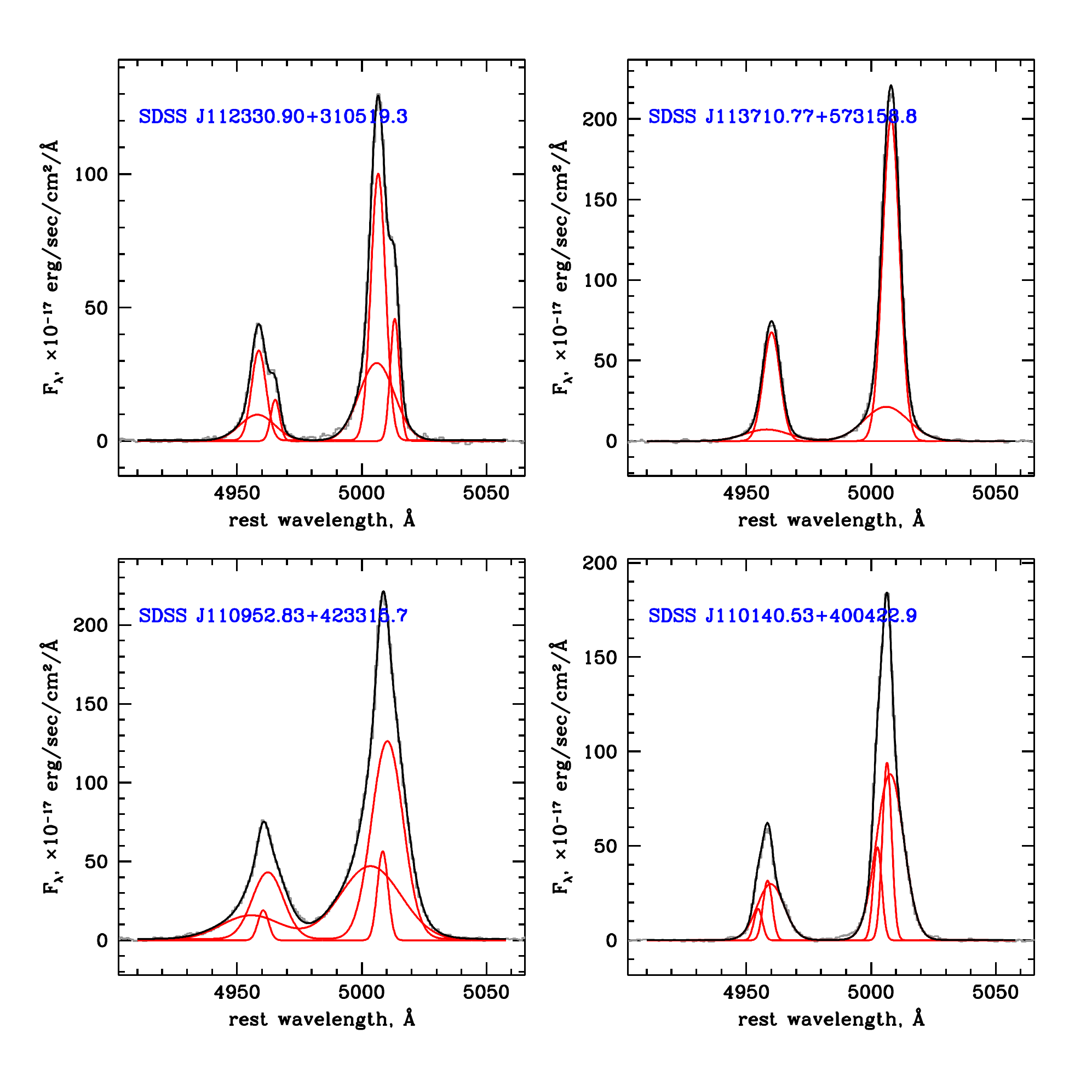}%
\includegraphics[scale=0.44, clip=true, trim=0cm 10cm 10.5cm 0.6cm]{picture_science33a}%
\includegraphics[scale=0.44, clip=true, trim=10cm 10cm 0.5cm 0.6cm]{picture_science33a}%
\includegraphics[scale=0.44, clip=true, trim=0cm 10cm 10.5cm 0.6cm]{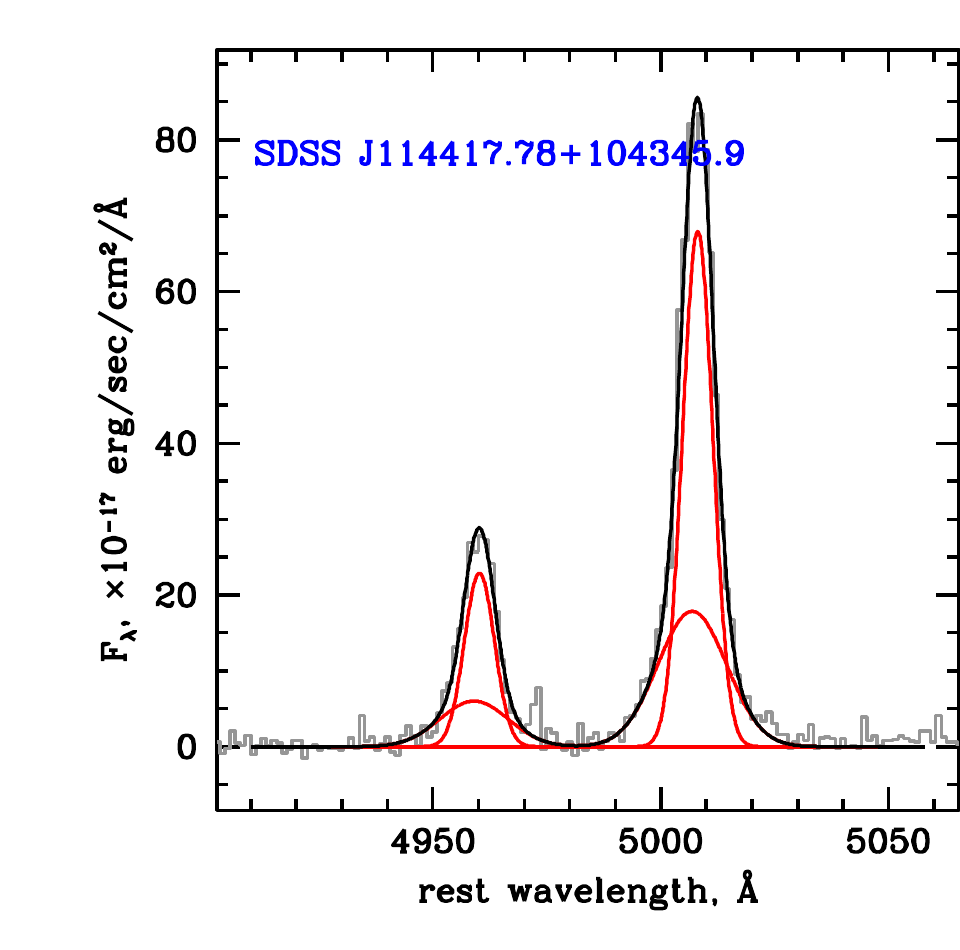}\\
\caption{Top: Four quasars that are spatially resolved in our A-array JVLA observations; from left to right, JVLA A configuration 6.0 GHz images at 0.495\arcsec\ resolution of three type 2 quasars and one type 1 quasar. The black star shows the optical coordinate of the source. The restoring beam is depicted as an ellipse on the lower left corner of each map and the scale is given on the right. Bottom: Continuum-subtracted SDSS spectra of the [OIII]$\lambda\lambda$4959,5007\AA\AA\ doublet (in grey) with multi-Gaussian decomposition in red and the total fitted profile in black (closely follows the grey data, so the data may be indistinguishable from the model).}
\label{fig:lowz_resolved}
\end{figure*}

\begin{figure}
\includegraphics[scale=0.18]{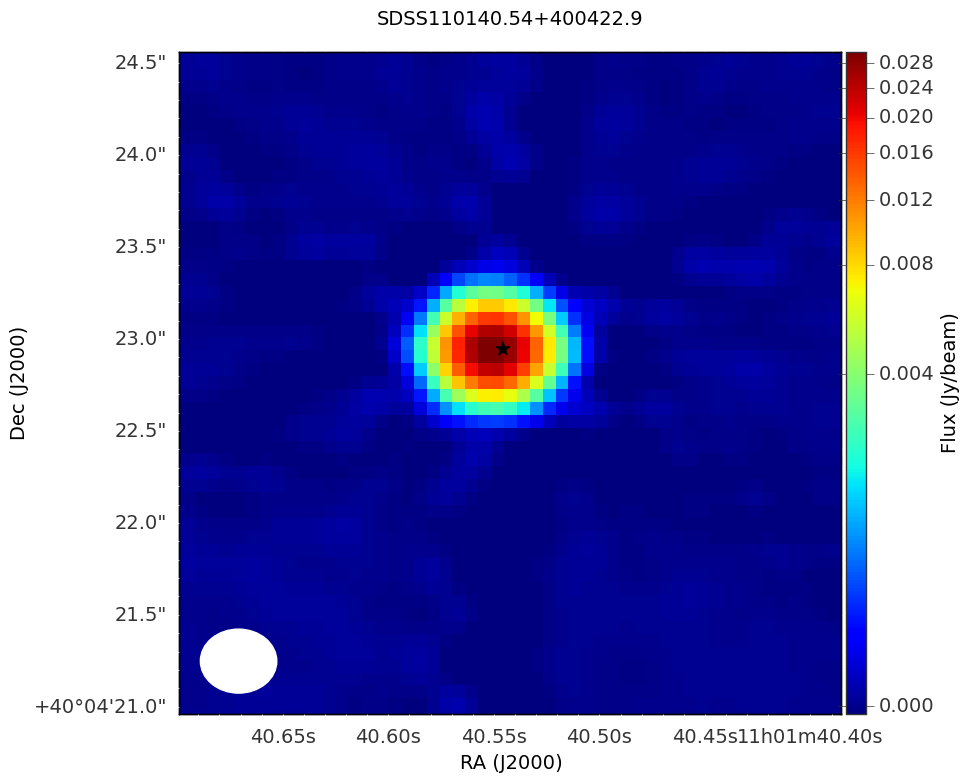}%
\includegraphics[scale=0.32, clip=true, trim=0cm 2.5cm 0cm 0cm]{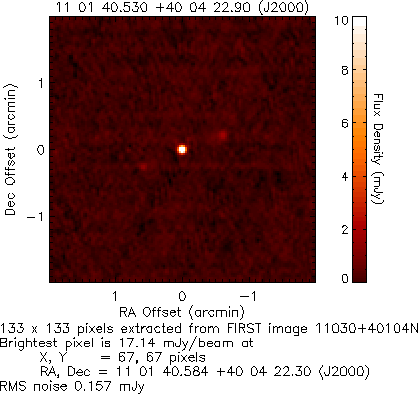}\\
\includegraphics[scale=0.44, clip=true, trim=10cm 0.6cm 0.5cm 10cm]{picture_science33a}\\
\caption{Top left and center: JVLA A configuration 6.0 GHz image of the one type 2 quasar SDSS~J1101+4004 which is a point radio source at resolution of 0.495\arcsec. Top right: the FIRST image on larger scales show that there are faint symmetric double lobes associated with the unresolved core. The projected physical distance between the cored and each of the lobes is $\sim$175 kpc. Bottom: the corresponding SDSS spectrum of [OIII] with the best multi-Gaussian fit.}
\label{fig:lowz_point}
\end{figure}

We resolve or marginally resolve four of the five sources we observed with the resolution of the A-array (0.5\arcsec; Figure \ref{fig:lowz_resolved}). The remaining source, SDSS~J1101+4004, is consistent with being point-like but it is almost certainly the core of a larger radio structure with lobes seen at $\sim 175$ kpc from the nucleus on either side (Figure \ref{fig:lowz_point}). This is strikingly different from our previous attempt to spatially resolve radio emission in radio-quiet type 2 quasars on somewhat larger scales: when we examined 19 objects that fell into the ``Stripe 82" region covered by 1.8\arcsec\ observations by \citet{Hodge2011}, none showed resolved structures \citep{Zakamska2014}. Therefore it appears that the typical scale of radio emission in radio-quiet quasars might be well-matched to the sizes of their host galaxies, as 1\arcsec$=5.4$ kpc at the median redshift of our five targets.

For the four objects that were previously detected in the FIRST survey, we can calculate the spectral index between the 1.4 GHz FIRST observations and our 6 GHz observations. The fifth object, SDSS~J1144+1043, is below the FIRST catalog threshold, but in the FIRST image centered on the SDSS position we see a 0.69 mJy/beam flux excess which we correct for CLEAN bias to yield an 0.94 mJy estimate for the 1.4 GHz flux. There is a significant time delay between the FIRST survey and our observations. Our spectral indices do not take into account the effects of variability on the measured fluxes, but unless much of the radio emission is concentrated in a pc-scale core, the variability on several year time scales is not a concern.  To measure spectral indices, we calculate the peak and integrated fluxes of our targets at 6.0 GHz using the CASA task {\sc imfit} to fit a 2D Gaussian to each source. In the case of SDSS~J1123+3105 we fit a gaussian to each component and label them left(l), right(r) and center(c) respectively. We use the larger of the peak and integrated flux as the radio flux. Spectral indices are determined between the FIRST fluxes and tapered 6.0 GHz fluxes with FWHM = 5\arcsec\ to match the resolution of the FIRST survey. The resulting spectral indices are listed in Table \ref{tab:lowz}. 

Four objects show steep spectral indices between $-0.5$ and $-1.0$, whereas SDSS~J1101+4004 -- the only point source in our sample and the only radio core associated with clear large-scale jets -- shows a flat spectral index of $0.4$. For SDSS~J1123+3105, the spectral index is calculated using the total flux from all three components at 6.0 GHz.  We also calculate a ``corrected" spectral index for the core assuming that both radio lobes have a steep spectrum of $\alpha = -0.7$ and subtracting the expected lobe radio flux at 1.4GHz from the FIRST flux to find a value of $\alpha_{\rm core} = -0.56$. Thus, in this object the core component also appears to have a steep index. 

Figures \ref{fig:lowz_resolved} and \ref{fig:lowz_point} also include our kinematic models for the [OIII]$\lambda$5007\AA\ emission of each quasar.  Each [OIII]$\lambda$5007\AA\ emission line is fit with one to three gaussian components depending on the reduced $\chi^2$ value of the fit (for more details see \citet{Zakamska2014}).  Note that in our model we assign no physical significance to each of multiple gaussian components but the presence in all of our objects of at least two gaussian components in their fit implies the existence of several structures moving at differing velocities.  There appears to be no statistical relation between the observed radio properties and the kinematics of the [OIII]$\lambda$5007\AA\ emission lines for our sample of five quasars, although with such a small sample size this is unsurprising.  We will explore in the next section several possible qualitative connections between the [OIII] kinematics and observed radio properties. 

\subsection{Discussion of low-redshift results}
\label{ssec:limp}

The radio emission observed in these five objects on kpc scales could be the result of several possibilities: (i) compact jets \citep{Leipski2006, Mullaney2013}, (ii) extended radio coronae \citep{Laor2008}, (iii) radiatively driven winds that produce radio emission as a biproduct \citep{Stocke1992, Zakamska2014, Nims2015}, and (iv) star formation in the host galaxy \citep{Kimball2011,Padovani2011,Condon2013}. 

We can rule out possibilities (ii) and (iv) in our sample.  Since four of the five targets are resolved on a few kpc scales and these same four objects also show steep spectral indices, this most likely rules out a radio corona as the dominant contributor to the radio emission: a putative radio corona would have scales of $\sim 10^5$ Schwarzschild radii -- i.e., parsec scales -- and because of its compactness would also be expected to have a flat spectral index.  The only object with both a compact radio source and a flat spectral index is SDSS~J1101+4004, but this is also the one object that shows large-scale jets, so we can be quite confident that we are seeing the core of a collimated jet in this source \citep{Blandford1979}.

In addition, the four type 2 quasars have infrared measures of star formation in the literature \citep{Zakamska2016a}. In all four cases the radio emission due to star formation is 0.6$-$1.7 dex below the observed emission, so although the objects are radio-quiet / radio-intermediate, star formation makes a negligible contribution, and the radio emission must be associated with the quasar activity. This leaves us with the possibility that the remaining four objects (excluding our large-scale radio jet in SDSS~J1101+4040) have radio emission powered either by compact jets or radiatively driven quasar winds. 

\subsubsection{Implications from radio morphology and [OIII] kinematics}

In three of our targets the spatially resolved emission is clearly directional. It is therefore tempting to classify them all as jets \citep{Leipski2006}. This assumption was recently called into question by \citet{Harrison2015} in a detailed observation of a nearby AGN called the ``Teacup". In this object, both the ionized gas emission and the radio emission are well resolved into a core and two lobe-like components. The lobe components are further identified in high-resolution observations as shells that are likely driven by an AGN-driven outflow. Although the source demonstrates a classical `core+lobes' morphology typically associated with collimated jets, \citet{Harrison2015} demonstrate that despite exquisite observations available for this source, what drives the outflow -- radiatively powered wind or collimated jet -- remains an unanswered question. 

This observation is particularly relevant for our study because the double-lobed morphology of our targets SDSS~J1123+3105 and SDSS~J1144+1043 is reminiscent of the ``Teacup", and the double-peaked kinematics of the [OIII] emission line we observe in these objects would also be a natural consequence of the two plowed shells, each with relatively small internal velocity dispersion, but each expanding with a significant overall velocity relative to the nucleus. Exactly this situation is seen in the ``Teacup" object where the two plowed shells are offset in velocity by about 330 km s$^{-1}$, while the two [OIII] components have FWHM of 250 and 325 km s$^{-1}$ respectively. In our source SDSS~J1123+3105, the two [OIII] components in our spectral fit are offset from each other by 330 km s$^{-1}$. If the outflow is propagating close to the plane of the sky -- as the radio morphology of our source might suggest -- then the physical velocity of the outflow is much higher than the observed velocity splitting because of the projection effects.  

In addition, type 1 quasar SDSS~J1144+1043 was previously observed using optical integral field unit spectroscopy \citep{Liu2014}. This object shows [OIII] emission with fairly low FWHM$\sim$200 km s$^{-1}$ (also consistent with a plowed shell) extending in the South-East direction (Figure 6 in \citet{Liu2014}, which is oriented with South at the top and East to the left). Therefore, we find that the axis of elongation for our radio image is qualitatively similar to the one seen in the ionized gas (see Figure \ref{fig:lowz_resolved}). However, the brighter radio lobe extends toward the North-West from the nucleus (and is thus co-spatial with the highest FWHM$\sim 900$ km s$^{-1}$ [OIII] component), whereas the brightest extended component of ionized gas as imaged in [OIII] is extended toward the South-East, in the direction of the fainter radio lobe. It is remarkable that these details are emerging from radio and integral-field unit observations of a quasar at $z=0.678$; for comparison, the ``Teacup'' is at $z=0.085$ though it does not yet allow us to differentiate between compact jets and radiatively driven winds as the source of radio emission in our quasar sample. 

\subsubsection{Implications from radio spectral indices}

It is possible that the three most extended sources -- SDSS~J1123+3105, SDSS~J1144+1043 and SDSS~J1137+5731 -- are similar to compact steep spectrum (CSS) or Gigahertz-peaked spectrum (GPS) objects \citep{Odea1998} in that they show compact radio emission and steep spectral indices. CSS and GPS objects are produced by powerful jets which strongly interact with the surrounding gas and produce very luminous radio lobes. While CSS and GPS sources are compact ($\la$ kpc) by the standards of radio galaxies ($\gg$ kpc), the lobe emission is extended enough that self-absorption is not important and thus as long as the lobe emission dominates over the cores, the indices are steep. Recent research at very high resolution that has been able to differentiate the morphology of GPS sources \citep[e.g.][]{Snellen2000,Taylor2000, Polatidis2002,Orienti2012,Orienti2014}, finds that many of them look morphologically similar to SDSS~J1123+3105 and SDSS~J1144+1043. The difference, however, is that GPS sources still have flat radio spectra in their core because on small scales the flux is dominated by the self-absorbed jet core \citep[e.g.][]{Snellen2000,Polatidis2002,Orienti2012,Orienti2014}. This contrasts with our observations of SDSS~J1123+3105 and SDSS~J1137+5731 where the central radio source is both dominant and has a steep spectrum. 

Therefore, the brightest compact parts of our targets are inconsistent with pure jet core emission. This implies that our objects either have unresolved steep-spectrum hotspots dominating the observed spectral index of their core component, or that we must seek an alternative explanation, such as quasar winds, for their radio flux. The presence both of extended radio lobes on several kpc scales and of young hot spots within the unresolved central component would imply that the quasar has had several episodes of strong interaction between the putative jet and the host interstellar medium -- the older episode resulting in the extended lobe emission and the newer episode resulting in the more compact emission in the unresolved core. Because the extended lobes expand with velocities 500$-$1000 km s$^{-1}$ as seen in their [OIII] emission over 5$-$10 kpc, their age must be about 10$^7$ years. Thus in order to explain our observations with hot spots from multiple episodes of AGN activity and jet / gas interactions, our quasars must have active episodes every 10$^7$ years and sufficient gas available to interact with the jet. Evidence for such episodes is also seen in a nearby quasar with a powerful outflow \citep[][Sun et al. 2016, in prep.]{Sun2014} and in large samples of FR IIs with no core emission \citep{Velzen2015}. 

Alternatively, there is no jet at all on any scale, and the radio emission is dominated by the particles accelerated in shocks which arise when a radiatively driven wind propagates through the galaxy. Because this emission is extended on scales $\gg 1$ pc, it is not affected by self-absorption the way the jet core is and its spectrum would be steep. In this scenario, the most extended double-lobed sources such as SDSS~J1123+3105 are produced when the wind ``breaks out'' of the galaxy and plows through the circumgalactic medium resulting in shells of ionized gas and radio emission. Such extended morphology can be produced by jet-driven and radiatively-driven outflows alike \citep{Harrison2015}. It is the characteristics of the central unresolved emission that might make it possible to distinguish between the two.  

SDSS~J1109+4231 is both a steep spectrum source and is marginally resolved but symmetric in our radio map. It also shows the most extreme [OIII]$\lambda$5007\AA\ kinematics from SDSS optical spectroscopy.  Thus, this object may be the best candidate for quasar-wind driven radio emission which could produce the observed spherically-symmetric outflow observed in the radio.  In this scenario the wind is in the early stages of development and has not yet engulfed the entire galaxy and thus the radio emission is relatively compact.  The wind has also not yet slowed down due to interactions with the galactic interstellar medium and therefore shows very high velocity components in [OIII]$\lambda$5007\AA\.  This scenario fits well, qualitatively, with what we observe in SDSS~J1109+4231. Spatial mapping of the [OIII]$\lambda$5007\AA\ emission on galactic scales and VLBI observations of the radio emission would help to confirm this scenario.

\subsubsection{Conclusions}

\citet{Lal2010} found that 64\% of a sample of $z\sim0.5$ Type 2 quasars from \citet{Zakamska2003} were flat-spectrum sources, with only 10\% being extended at 0.8\arcsec\ resolution.  Therefore, our results are quite different from those of \citet{Lal2010}. As far as we know, the only difference in the selection of the two samples is the [OIII]$\lambda$5007\AA\ luminosity, with \citet{Lal2010} objects being about an order of magnitude less [OIII]-luminous than the ones discussed here (Figure \ref{pic:radio_power}). 

If high bolometric luminosity (and with it, stronger [OIII]$\lambda$5007\AA\ ) is a necessary condition for the existence of powerful quasar winds \citep{Zakamska2016b}, the difference between our observations and those of \citet{Lal2010} may in fact indicate that we are probing the transition between the low luminosity regime where large-scale winds and associated radio emission do not exist and the high luminosity regime where such an effect is prevalent.  Some evidence for such a luminosity threshold is found in observations of molecular outflows \citep{Veilleux2013}. In theoretical models \citep{Zubovas2012}, the threshold quasar luminosity might arise because a more powerful quasar-driven wind is necessary to push a given amount of material out of a galactic potential. Our assessments are currently speculative with such a small number of objects but hopefully a future program at a similar resolution and sensitivity would allow us to perform a more thorough investigation into the connection between the radio emission in radio-quiet quasars and [OIII]$\lambda$5007\AA\ kinematics.  In addition, VLBI observations of our objects with higher radio flux would allow us to probe to even higher resolution and perhaps differentiate these scenarios. Even from our small sample, it can be seen that the combination of radio morphology, power and spectral index can be a powerful tool to differentiate the origin of the radio emission in radio-quiet quasars.  

\begin{deluxetable*}{l|r|r|r|r|r|r|r|r|r|l}
\tablecolumns{11}
\tablecaption{Low Redshift Sample Properties\label{tab:lowz}}
\tablehead{
Source name, & SDSS & $F_{\nu}^{\rm peak}$, & $F_{\nu}^{\rm int}$, & $F_{\nu}^{\rm rms}$, & $F_{\nu}^{\rm peak}$, & $F_{\nu}^{\rm int}$, & $F_{\nu}^{\rm rms}$, & spectral & $L$[OIII] & RL? \\
SDSS coordinates & redshift & 6 GHz & 6 GHz  & 6 GHz  & 1.4 GHz  & 1.4 GHz  & 1.4 GHz & index $\alpha$ & & }
\startdata
SDSS~J110150.53+400422.9 & 0.4569 &32.3&32.2&0.015&17.85&17.38&0.142&0.40&9.71&RI\\
SDSS~J110952.82+423315.7 & 0.2612 &2.8&3.6&0.0157&16.21&16.96&0.135&-0.75&9.37&RQ\\
SDSS~J112330.90+310519.3(c) & 0.3097 &6.9&6.9&0.00938&17.27&19.89&0.122&-0.73&9.15& RI\\
SDSS~J112330.90+310519.3(l) & & 1.3& 2.9& 0.00938& & & & & & \\
SDSS~J112330.90+310519.3(r) & & 0.22& 0.43& 0.00938& & & & & & \\
SDSS~J113710.77+310558.8 &  0.3953 & 0.29& 0.36& 0.00696& 2.35& 2.14& 0.143& -1.0& 9.6& RQ\\
SDSS~J114417.78+104345.9 &  0.6783 & 0.052& 0.45& 0.00836& 0.94& & 0.154& -0.51& 9.67& RQ\\
\enddata
\tablecomments{Peak fluxes and rms values are in mJy/beam and integrated fluxes are in mJy. [OIII] luminosities are given as $\log$($L$[OIII]/$L_{\odot}$). Designation as radio-loud (RL), radio-intermediate (RI) or radio-quiet (RQ) is based on the position of the object in the $L$[OIII] vs 6 GHz luminosity space \citep{Xu1999}.}
\end{deluxetable*}

\section{Source Populations}
\label{sec:sourcepop}

In this section we describe our efforts to determine the nature of the sources identified in our radio fields using publicly available multi-wavelength surveys.  We compare our results to those of recent deep radio surveys. Automatic source identification is accomplished using {\sl Aegean} \citep{Hancock2012}. This source finding algorithm assumes a compact source structure allowing it to fit multiple components (as determined by a curvature map) to each island of pixels identified in the algorithm.  This enables {\sl Aegean} to identify both faint sources close to the detection limit and sources within an island of pixels containing multiple components and it produces both a reliable and complete catalog. We ran {\sl Aegean} on each of our fields to produce a catalog of all sources present at a detection limit of 6$\sigma$.  This limit is chosen because there appeared to be a significant percentage (42\%) of false positives around 5$\sigma$.  All sources from this catalog are included in the analysis unless it is clear that the program had mis-identified residual flux as a source.  We excluded such sources by eye. 

\begin{figure}
\includegraphics[scale=0.5,trim=90 350 50 75,clip=true]{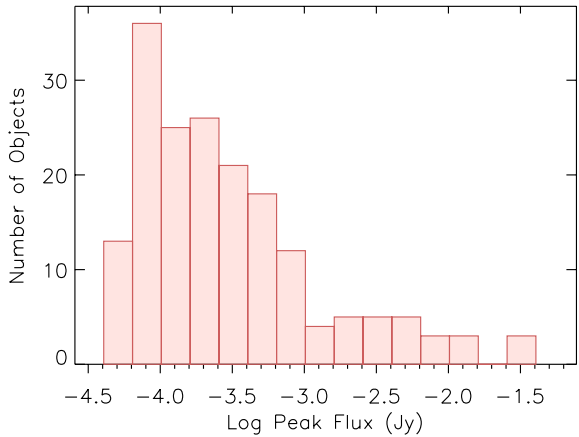}
\caption{Distribution of 6 GHz radio flux densities for all sources identified in our twenty radio fields by {\sl Aegean}.}
\label{pic:radio_hist}
\end{figure}

In total we identify 179 sources in our fields using {\sl Aegean}, with fluxes between $\sim 40$ \ujy\ and $\sim 40$ mJy and a median flux of 0.23 mJy.  Their peak flux densities are shown in Figure \ref{pic:radio_hist}. The full table of all sources is available online with the first 20 rows included here as Table \ref{tab:sample}. With the 6 GHz-selected source catalog in hand, we use {\sl Aegean} in forced measurement mode to calculate the flux at our source locations in matching L-band data for fields in Program 13B-382 where it is available and in any tapered fields created for comparison (see \S~\ref{ssec:slope}). 

{\sl Aegean} returns both a point source flux and an integrated flux for all islands and for each component within a given island. Unfortunately, the current version, as of February 14 2016, does not correctly calculate the integrated flux of identified islands, though the integrated flux of components is correct.  Thus, unless otherwise indicated, we choose as our sources {\sl Aegean} islands, but only report and use their peak flux in our measurements. This means we may be underestimating the true radio flux in our resolved sources. Six of our sources identified with {\sl Aegean}, two of which are our targets, upon visual inspection are clearly multiple components of an extended source with lobes or other features.  All six sources are identified as single sources in FIRST. 

We cross-matched all of our sources identified using {\sl Aegean} with the FIRST catalog. In the case of our B-configuration fields we have L-band data at a similar resolution to FIRST which provides a direct check on our measured fluxes. We create a test catalog using {\sl Aegean} to identify sources in our L-band fields and then cross-match this catalog with FIRST using a matching radius of $<$4\arcsec.  For sources where FIRST matches existed, typically 97\% of sources above 2 mJy and 25\% of sources below 2 mJy, we found a mean flux ratio of flux$_{\rm Aegean}$/flux$_{\rm FIRST} = 0.83$ and a standard deviation of 0.21. We are thus confident that {\sl Aegean} is capturing most of the radio flux in our fields; the remaining discrepancy is accounted for by the CLEAN bias (taken into account in the FIRST catalog but not in our images) and the missed flux in spatially resolved objects.

To verify that the faint sources we detect using {\sl Aegean} are real, we stack FIRST cut-outs of all of sources not individually detected in FIRST (121 sources). The resulting FIRST stack at 1.4 GHz has a median flux of 0.16 mJy, 6 times below the FIRST detection limit, and a mean flux of 0.18 mJy. (0.25 mJy when accounting for Snapshot bias).  For comparison, the mean and the median flux of the same sources measured by {\sl Aegean} at 6 GHz is 0.22 mJy and 0.14 mJy respectively.  Assuming a spectral index of $\alpha = -0.7$ our extrapolated 1.4 GHz flux is a factor of two greater than the FIRST flux measured in our stacked image or it implies a flat spectral index of $\alpha = -0.1$. We have reason to suspect a flatter spectral index would be more appropriate for our 6 GHz-selected catalog (see \S~\ref{ssec:slope}), decreasing this discrepancy.

\subsection{Resolved Sources}
\label{ssec:resolved}

There are several accepted methods in the literature to determine the percentage of resolved sources in a survey.  In particular, the obvious definition is to check if a source's FWHM is greater than the beamsize taking into account the effects of bandwidth smearing as a function of position and depth.  Using this technique, the FIRST survey found that $\sim 20\%$ of their sources are resolved with a beam FWHM of 5.4\arcsec\ and \citet{Hodge2011} found that $\sim 60\%$ of their sources are resolved in their survey of Stripe 82 with a beam FWHM of $\sim 2.3$\arcsec.  We find, for the gaussian components that make up our source islands, not taking into account the effects of beam smearing, that $50\%$ of the components we observed with the highest resolution (A-array) had a FWHM $> 0.495$\arcsec\, which is the approximate beamsize at the center of the image given natural weighting. 

Alternatively, surveys such as VLA-COSMOS \citep{Schinnerer2004} and E-CDFS \citep{Miller2013} compared the ratio of total integrated flux/peak flux to peak flux/rms noise. They take the definition of ``resolved" to be total/peak$ > 1+$A/(peak/rms)$^3$ where A is some constant set such that some large percentage of sources ($\gtrsim 95 \%$) fall above the function flipped over the axis total/peak = 1.  In good agreement with our previous value, we find that 52.4\% of the components that make up our sources in our fields observed with the highest resolution A-array are resolved according to this definition, when we set $A=100$ (which bounds 98\% of our sources).  In addition, according to this definition we find that 52.2\% of the components that make up our sources observed in the B-array configuration (approximate beamsize of 1.3\arcsec\ with natural weighting) are resolved. In comparison, the VLA-COSMOS survey found that $44\%$ of their sources are resolved with a beam FWHM of $\sim 1.5$\arcsec\ and the E-CDFS survey found $14\%$ of their sources are resolved with a beam FWHM of $\sim 2.2$\arcsec\ 
Thus, our observations seem consistent with the overall trend, unlike for our smaller sample of low redshift type 2s where we appear to see a large increase in the fraction of resolved sources at a specific resolution.   

\subsection{Source Counts}

Our first task is to verify that we recover the faint radio sources at source number densities comparable to those found in other deep radio surveys. We use the 6 GHz catalog from both the A-array and B-array observations. The effective area of our observations varies as a function of flux because of primary beam attenuation. For every source flux, we use the primary beam shape to estimate the distance from the center of the field at which such source flux would become non-detected and use this distance to calculate the effective area of the survey at this flux. We then calculate the number of sources detected per unit flux and per unit effective area of the survey, which is well fit by a power-law between 0.1 and 10 mJy:
\begin{eqnarray}
    \log\left(F_{\nu}^{5/2}\frac{{\rm d}N}{{\rm d}F_{\nu}},{\rm Jy}^{3/2}\mbox{ sr}^{-1}\right)=\nonumber \\
    \left(3.0\pm0.2\right)+\left(0.8\pm0.1\right)\log\left(F_{\nu},{\rm Jy}\right).
\end{eqnarray}
(The errors include only the error in the fit, but not the cosmic variance. This should be considered only an estimate as the same sensitivity curve is assumed for all the fields.)

This is in excellent agreement with the compilation of source counts presented by \citet{Massardi2010} at 5 GHz (which we correct to 6 GHz using a spectral index of $-$0.7). According to their population synthesis models, 0.1 mJy at 5 GHz is precisely the flux above which AGN should dominate the radio population and below which star-forming galaxies are expected to dominate. Unfortunately, our survey becomes incomplete below 0.1 mJy, but in what follows we attempt to elucidate the nature of the bulk of our detections between 0.1 and 10 mJy for comparison with such population synthesis models. 

\subsection{Spectral Indices}
\label{ssec:slope}

For our high-redshift fields in Program 13B-382, spectral indices are determined by combining the C and L-band data.  For a proper comparison, new C-band maps are created at the resolution of the L-band using {\sl CLEAN} with a uv-taper FWHM of 6.45\arcsec\ (the approximate resolution of the L-band in B-configuration if imaged with natural weighting). We then force {\sl Aegean} to calculate the peak flux in our L-band maps and tapered C-band maps using locations provided from our full resolution C-band maps. A spectral index is then calculated using our L-band data at 1.4 GHz (whenever the flux from forced photometry is positive) and our tapered C-band data at 6.0 GHz.  We compare the measured C-band flux at full resolution to the calculated flux using our uv-tapered map and identify a sample of point sources for which the full resolution map and lower resolution map have good agreement (the measured flux ratio is smaller than 5) though we note that the median slope of our full sample is actually steeper by only -0.04 with a similar standard deviation.  We find a flat median slope of $\alpha_{med} = -0.19$ with a standard deviation of 0.66.

Although there are some hints of spectral index flattening toward lower radio fluxes \citep{Gralla2014,Huynh2015}, such values are surprising. To understand this observation we conduct a simulation which closely mimics our observations. We draw sources from the radio flux distribution of \citet{Massardi2010}, ``observe'' them including the effects of the primary beam on the detection sensitivity both at 1.4 and 6 GHz, include CLEAN bias, typical observational errors and our catalog strategy, in which the sources are selected at 6 GHz and then forced photometry is obtained at 1.4 GHz. We put in a Gaussian distribution of spectral indices with mean $\alpha_{\rm im}=-0.7$ and standard deviation $\sigma_{\alpha}=0.5$, and then use the simulation to determine the mean and the standard deviation of the recovered spectral index as a function of 6.0 GHz flux for our survey. 

We find much flatter output spectral indices, $\alpha_{\rm out}\simeq-0.2$, similar to what we observe, with overall trend $\alpha_{\rm out}\simeq 1.4\sigma_{\alpha}-0.9$.  The most important reason for this strong spectral index bias is that our catalog is selected at 6 GHz and we dig well into the noise at 1.4 GHz to recover objects with the flattest spectral indices, whereas the objects with the steepest indices and faint (but detectable) 1.4 GHz fluxes would never make it into our catalog as they are not detected at 6 GHz. Additional complications include the difference in the shape of the primary beam, whose size is $\propto 1/\nu$. As a result the 1.4 GHz data is not as sensitive as our 6.0 GHz data in the center of the field, so we would have difficulty identifying flat spectrum objects among the faint sources which can only be identified near the field center. As we step away from the center of the field, the shape of the primary beam makes 1.4 GHz data more sensitive than the 6 GHz data, but the flux limit for the detection is now much higher. As a result, there is a slight steepening of the observed indices at the faintest fluxes in the simulation which is also seen in the data. These effects are a result of the fact that our survey is not a mosaic of a particular field and so our sensitivity varies across each field. These effects swamp much of the intrinsic effects we would hope to observe as trends in our spectral index as it evolves with the changing distribution of source type. 

\subsection{Cross-matching with the SDSS data}
\label{ssec:sdss}

\begin{figure}
\includegraphics[scale=0.45,trim=20 45 0 250,clip=true]{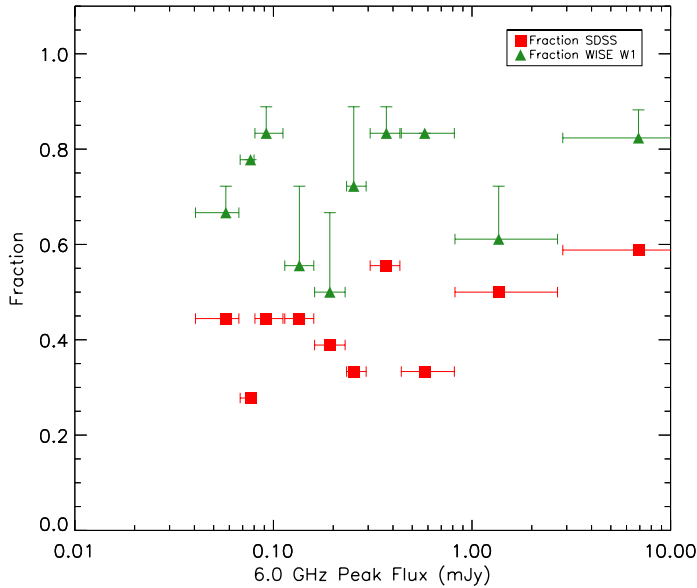}
\caption{Fraction of detections as a function of 6.0 GHz peak flux.  We divided the sample into 10 bins with an equal number of sources in each bin and placed the data point at the median flux in each bin.  The range of fluxes in each bin is represented by the horizontal error bars.  The green triangles show the fraction of sources in each bin with a WISE detection within 3\arcsec\ and a S/N ratio $> 3$ in W1. We plot an upper limit for each bin by including all WISE sources detected within 6\arcsec\ with a S/N ratio $>3$ in W1.  The red squares show the fraction of sources in each bin with an SDSS detection within 1\arcsec.}
\label{pic:fraction_coverage}
\end{figure}

We cross-match our 179 sources with the SDSS data release 12 \citep{Alam2015} photometric and the spectroscopic catalogs, within 1\arcsec. We find 47 photometric matches and 24 spectroscopic matches (excluding our original targets), for a total recovery fraction of 41\%. Out of the sources matched in SDSS, 91\% have either a photometric redshift estimate \citep{Csabai2007} or a clear measurement from SDSS spectroscopy (Table \ref{tab:sample}). The matching rate as a function of radio flux is shown in Figure \ref{pic:fraction_coverage}.

We examine the 24 spectra as well as their classifications from the Portsmouth group \citep{Thomas2013}. We use the spectroscopic classifications from the Portsmouth group in all cases except two for which both human classifiers agreed with each other but disagreed with the pipeline.  Our spectroscopic identifications are seven AGN (including low-redshift Seyferts, high-redshift quasars, one serendipitous $z=0.44$ type 2 quasar and one blazar), two star-forming galaxies with redshifts 0.088 and 0.073 respectively, and ten absorption-line galaxies. In addition, there are four LINERs and one composite object which is an absorption line galaxy that shows signs of AGN activity.  The median radio flux at 6 GHz of this sample is 0.39 mJy, while the mean flux is 3.4 mJy, dominated by the seven brighter sources with flux $> 1$mJy. Of the spectroscopic matches identified as AGN, 33\% are above the median radio flux of our full sample. One of our spectroscopic SF galaxies falls above the median peak flux of our total radio sample while one falls below.  

The photometric redshift estimator of \citet{Csabai2007} also provides optical classification based on the best fit to the spectroscopic templates of \citet{Dobos2012}.  We create three categories for optical classifications: AGN, SF galaxies, and passive galaxies. We categorize any object showing signs of AGN activity in the optical (including LINERs from the spectroscopic sample) as AGN. Only objects without any sign of AGN activity are placed in the SF category. Our final category, passive galaxies, showed no sign of either AGN or star-formation activity in spectroscopy or in the photometric template used in optical classification.

\begin{figure}
\includegraphics[scale=0.5,trim=70 350 0 100,clip=true]{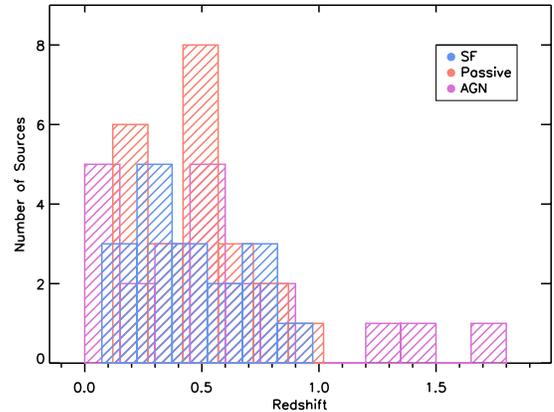}
\caption{Redshift distribution of all of our objects with either a photometric or spectroscopic redshift in SDSS. The different subsamples are classified according to \S~\ref{ssec:sdss}.}
\label{pic:photoz}
\end{figure}

\begin{figure}
\includegraphics[scale=0.75,trim=120 30 200 150,clip=true]{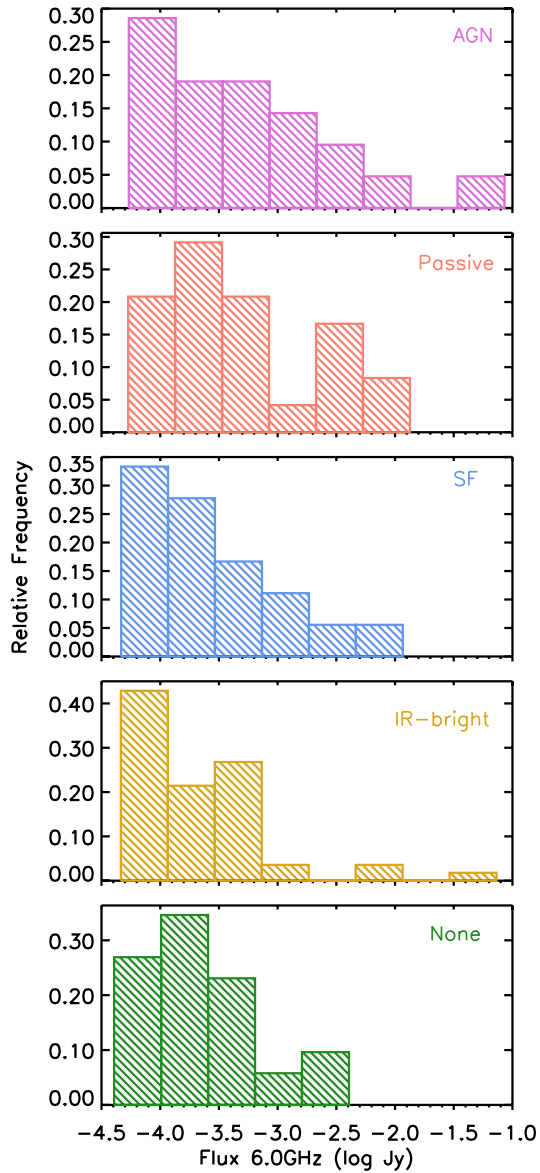}
\caption{Distribution of radio flux densities at 6.0 GHz of all of our objects with WISE and / or SDSS counterparts. The different subsamples are classified according to \S~\ref{ssec:wise} and \S~\ref{ssec:sdss}.}
\label{pic:compare_radioflux_type}
\end{figure}

Using this classification scheme, of the 77 sources with SDSS spectroscopy or photometry, 22\% of our sources are classified as SF galaxies, 29\% are classified as AGN and 30\% are classified as passive galaxies.  The remainder are either target sources (8\%), or did not have a photo-z so remain unclassified in all future analysis.  We show the full redshift distribution of our sources in figure \ref{pic:photoz}. The mean redshift of the sample is $z = 0.48$ though this represents only those sources which had a match in SDSS -- the missing sources with no counterparts in the optical are likely to be at a higher redshift. In addition, in Figure \ref{pic:compare_radioflux_type} we show the radio flux distribution of our samples, including an additional category for both unclassified objects (no detection in SDSS or WISE, or an SDSS detection without a redshift and classification) as well as IR-bright objects (those detected in WISE W1 but not SDSS, see Section \ref{ssec:wise}). 

Interestingly, a significant fraction of our sources are identified as passive or absorption line galaxies -- showing neither star formation or AGN activity in the optical. This population is most plausibly identified with the ``low-excitation" or ``radio-mode" AGN of \citet{Best2012}. Such a population would be mis-identified as star-forming galaxies in surveys that assume all objects not classified as AGN are star-forming galaxies and require optical, infrared or X-ray signatures for AGN classification. For example, \citet{Bonzini2013} find a high percentage (60\%) of SF galaxies among their faint radio sources, probably because they require X-ray or mid-infrared signatures of AGN activity to be present in order to classify their sources as AGN. Similarly, \citet{Smolcic2008} require either emission line activity or X-ray activity to classify sources as AGN candidates, and it is not clear how passive galaxies would be classified in their scheme. 

\subsection{Cross-matching with WISE data}
\label{ssec:wise}

We use the All-WISE catalog of the Wide-Field Infrared Survey Explorer \citep[WISE;][]{Wright2010} to investigate the mid-infrared properties of the faint radio population detected in our observations. Of the 179 sources individually detected in all 20 fields in the C-band, we find 128 WISE matches within 3\arcsec\ and 143 matches within 6\arcsec.  Because there is not a well-pronounced minimum in the distribution of the matching distance, it is not clear what matching radius is more appropriate. We consider the matches within 3\arcsec\ to be robust (listed in Table \ref{tab:sample}) and the other ones to be tentative.

In Figure \ref{pic:fraction_coverage} we show the fraction of sources with robust WISE matches as a function of the measured radio flux. The upper limit shows how the fraction of detections in each bin would change if we instead used a matching radius of 6\arcsec. The overall detection rate remains high even down to our flux limit, with a detection rate of 67\% of sources in W1 with fluxes between 40.5 and 67 \ujy. Essentially all SDSS matches are also matched in WISE, but the reverse is not true: we define a category of objects, henceforth referred to as ``IR-bright", for which there is a WISE detection in W1 but no SDSS detection. 

\begin{figure}
\includegraphics[scale=0.5,trim=60 350 10 100,clip=true]{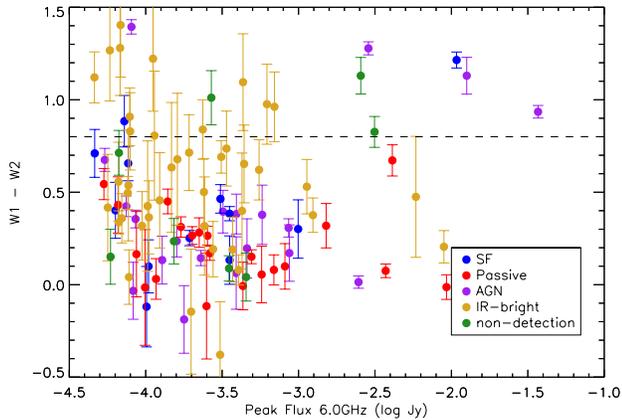}
\caption{WISE colors of the infrared-detected counterparts to radio sources, classified as described in \S~\ref{ssec:wise} and \S~\ref{ssec:sdss}. The Rayleigh-Jeans tail of a stellar population has W1$-$W2$\simeq 0$ mag because the WISE photometric system is Vega-based. Objects with significant hot dust contribution (such as luminous AGN) typically have W1$-$W2$>0.8$ mag. While this is a sufficient condition for AGN selection, it is not a necessary one, as the spectral energy distribution of dust can start rising at much longer wavelengths depending on the geometry of obscuration, and such sources will appear to have W1$-$W2 colors typical of a normal stellar population.}
\label{pic:wise}
\end{figure}

Interestingly, the median W1$-$W2 colour for these objects has a value 0.25 mag higher than the SDSS-detected sample (Figure \ref{pic:wise}). Of particular interest is the condition W1$-$W2$>0.8$ mag, which is often used to identify infrared-luminous AGN \citep{Assef2010, Stern2012}. While this condition does not select all AGN (an obscured AGN spectral energy distribution can be dominated by the host galaxy with W1$-$W2$<0.8$ mag if the warm dust is not visible from the observer's direction, \citealt{Mateos2013}), samples selected by this cut have low contamination rates. 29\% of objects in our IR-bright sample lie above this threshold, compared to only 18\% of SDSS-detected objects. Combined with non-detection in the SDSS, this observation likely indicates that the IR-bright sample may be dominated by obscured AGN. 

Using the IPAC Infrared Science Archive, we extract and stack WISE exposures of individually non-detected faint radio sources, centered on the {\sl Aegean}-derived positions.  We find no detection in any of our stacks.  The 2-$\sigma$ brightness upper limit in our W1 stack corresponds to a flux of $\approx6~\mu$Jy.

\subsection{Star Formation Rates}
\label{ssec:sfr}

We use the 6.0 GHz radio flux and SDSS redshift of our sources, where available, to estimate a star formation rate (SFR). SFRs are estimated using equation (6) of \citet{Bell2003}, where we can calculate the rest-frame 1.4 GHz luminosity by k-correcting as in our equation (\ref{eq_lum}), with the resulting star formation rate:

\begin{equation}
\psi(\rm M_{\odot}\mbox{ yr}^{-1}) = 3.94\times 10^{-38} \left(\frac{\nu L_{\nu}{\rm [1.4\, GHz]}}{\mbox{ erg s}^{-1}}\right).
\end{equation}

\begin{figure}
\includegraphics[scale=0.5,trim=70 350 0 100,clip=true]{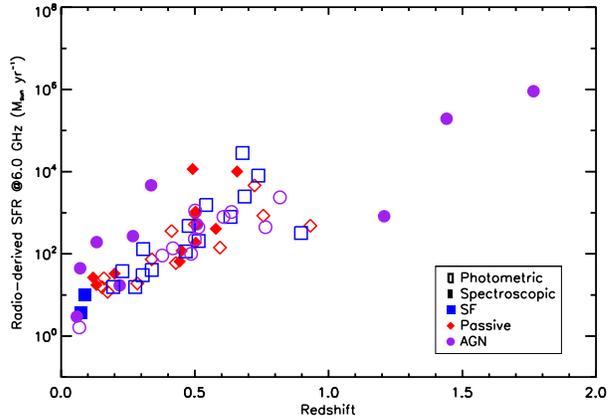}
\caption{SFR as a function of redshift for all of our sources with a match in the SDSS spectroscopic or photometric sample. Sources are identified as AGN, SF or passive galaxies based on either spectroscopy or templates used in photometric redshift determinations. SFRs of $>10^4$ are unlikely even in the high redshift universe implying that some of the measured radio flux must be due to a different source.}
\label{pic:compare_sfr}
\end{figure}

Calculated SFRs are shown in figure \ref{pic:compare_sfr}, where we have assumed $\alpha=-0.7$ (radio luminosities and SFRs smaller by a factor of 2 would be inferred if $\alpha=-0.2$ is assumed instead).  A star-formation rate greater than 1000 $M_{\odot}$ yr$^{-1}$ would be comparable to some of the most active sub-mm galaxies in the universe \citep{Casey2014} making our calculated SFRs, especially for most objects identified as AGN, improbable.

We can test whether our radio SFR estimates agree with the observed WISE fluxes. At a redshift of $\sim 0.5$, the strong polycyclic aromatic hydrocarbon emission feature at 7.7\micron\ falls into the center of the W3 band, so W3 fluxes would be a sensitive measure of extreme star formation at those redshifts. We take seven objects with measured W3 fluxes (the rest are upper limits) and use SF templates from \citet{Mullaney2011}, placed at the redshift of each object, convolved with a W3 filter curve and scaled to the observed W3 flux to calculate the total luminosity of star formation which we then convert to a star-formation rate \citep{Bell2003}.  We find star formation rates ranging between 2 and 25 $M_{\odot}$ yr$^{-1}$ (with good agreement between the W3-based and the radio-based star formation rates for the two spectroscopic SF galaxies), with only one object (photometric redshift $z\approx0.7$) at 200 $M_{\odot}$ yr$^{-1}$. We find no star formation rates as high as would be necessary to explain the radio fluxes we see in this subsample. 

This leads us to conclude that the majority of the objects classified by the SDSS photometric pipeline as SF galaxies in our sample are in fact AGN. With only five optical bands and often at the limit of the optical survey, distinguishing between an SF galaxy and an AGN is very difficult. Both types of objects show emission lines and blue continua of varying strength. We add additional information with our radio detection.  The two spectroscopically confirmed SF galaxies (both at $z<0.1$ and with SF rates of 2$-$5 $M_{\odot}$ yr$^{-1}$) are robust and consistent with all observations. 

\subsection{Summary and discussion of the study of faint radio sources}

We identify 179 radio sources in the C-band observations. 41\% of these have optical matches in the SDSS (magnitude limit of $i_{AB}=21.3$) , and an additional 33\% have matches in the WISE survey, but not in the SDSS. Our matching fraction is somewhat dependent on the radio flux, with a higher matching fraction at higher fluxes. \citet{Smolcic2008} investigate the nature of the radio sources in the VLA-COSMOS survey using follow-up optical data down to the depth of $i_{AB}=26$ mag. They find optical counterparts to 65\% of their sources. Because we are sensitive to faint radio sources only over a small fraction of our observing area, the flux distributions of radio sources in our mini-survey are biased toward higher values than that of \citet{Smolcic2008}, and therefore our optical detection fraction of 40\% is higher than what would be naively predicted from the flux distribution of matches in their survey. 

We classify our detected counterparts into four roughly equal categories based on their optical and infrared properties: AGN, passive galaxies (which have no signs of AGN activity other than the radio source), star-forming galaxies and IR-bright sources. Star-forming galaxies are primarily identified as such by the SDSS photometric redshifts pipeline \citep{Csabai2007}, with only two star-forming galaxies spectroscopically confirmed. Estimating star formation rates from the measured radio fluxes and the nominal photometric redshifts, we find that at 73\% of our star-forming galaxy classifications have radio fluxes that imply star formation rates $>100$ M$_{\odot}$yr$^{-1}$ while 33\% of our star-forming galaxies have radio fluxes that imply star formation rates $>500$ M$_{\odot}$yr$^{-1}$ which is implausibly high especially given the star formation rate estimated from the W3 flux which covers the 7.7$\mu$m PAH feature at these redshifts. Thus, these objects are probably mis-identified AGN which are difficult to distinguish from star-forming galaxies on the basis of five photometric measurements of the SDSS, especially close to the magnitude limit of the survey. 

Because we have a set of pointed observations rather than a mosaic, our survey depth is not uniform and is strongly affected by the frequency-dependent primary beam. Correcting for this effect, we estimate that our number counts in the C-band are consistent with the previous compilations by \citet{Massardi2010}. These authors predict from the population synthesis models that the $<0.1$ mJy population should be dominated by star-forming galaxies. The vast majority of the optical and infrared counterparts that we find are not star-forming galaxies. Thus either star-forming galaxies dominate at even lower fluxes than our survey probes or they all fall below the flux limit of the SDSS survey. 

\section{Interesting sources}
\label{sec:interesting}

\subsection{Dual AGN}

\begin{figure}
\includegraphics[scale=0.18]{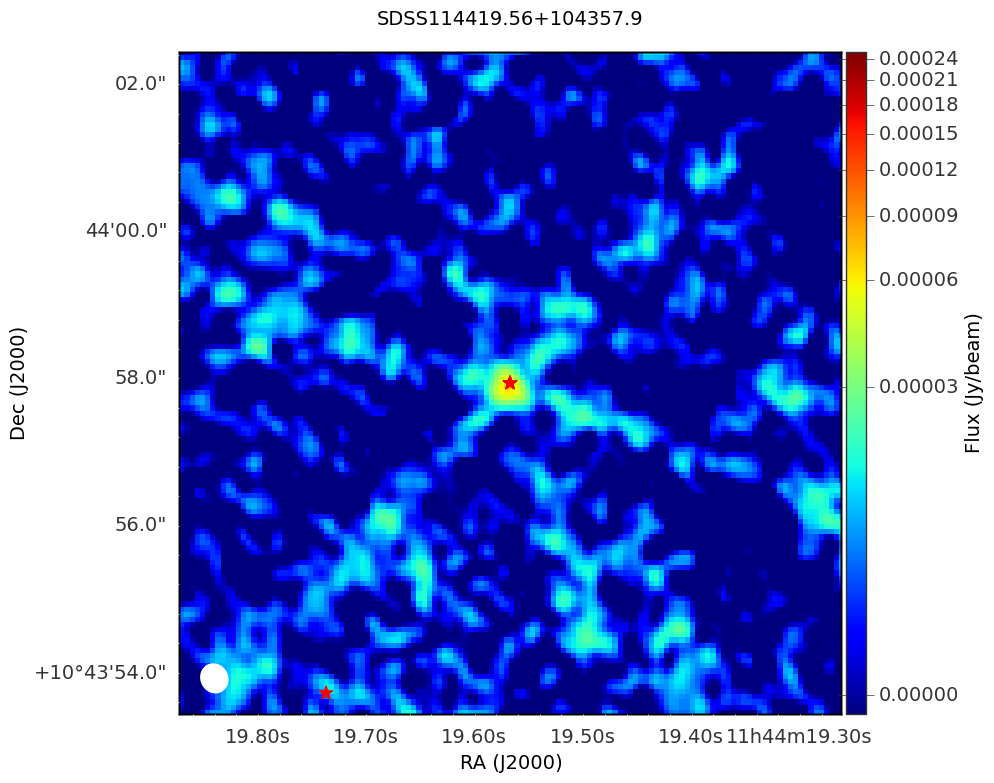}%
\includegraphics[scale=0.4, clip=true, trim=10cm 10cm 0.5cm 0.6cm]{picture_boss4}\\
\caption{JVLA image of the pair of active galaxies SDSS~J114419.74+104353.7 (left bottom red star marks optical position) and SDSS~J114419.56+104357.9 (center, detected radio source, optical position marked with red star). The first object is an optically identified type 2 quasar which is not detected in the radio, whose [OIII]$\lambda\lambda$4959,5007\AA\AA\ emission is shown in the right panel. The second object is detected in the radio, but shows a pure absorption-line spectrum in the SDSS with no optical signs of AGN activity.}
\label{pic:binary}
\end{figure}

In the A-array observation of type 1 quasar SDSS~J114417.78+104345.9 (which is at redshift $z=0.678$) we found two objects with SDSS optical spectroscopy unrelated to our target (Figure \ref{pic:binary}). Both sources are at $z=0.444$, separated by about 4\arcsec, or 23 kpc. The northern source is an absorption line galaxy with no optical or IR signs of AGN activity (it is detected in WISE with W1$-$W2 = 0.34 mag), and it is also a 63.2 $\mu$Jy radio source, with a corresponding luminosity $\nu L_{\nu}$[1.4 GHz]$=1.57 \times 10^{39}$ erg s$^{-1}$. Its estimated mass from SDSS is $10^{10.9}$ M$_{\odot}$. The southern source is an optically identified type 2 quasar, but without a radio counterpart and an estimated mass of $10^{11.5}$ M$_{\odot}$.  The measured FWHM([OIII]) from SDSS of the type 2 quasar implies an expected radio luminosity of $\nu L_{\nu}$[1.4 GHz]$=8.5\times 10^{39}$ erg s$^{-1}$ according to our equation (\ref{eq:fwhm}), which should be easily detectable by our survey.  It is not possible to tell whether the two galaxies are in fact physically interacting. 

\subsection{Objects with interesting structure in radio images}

\begin{figure}
\includegraphics[scale=0.17]{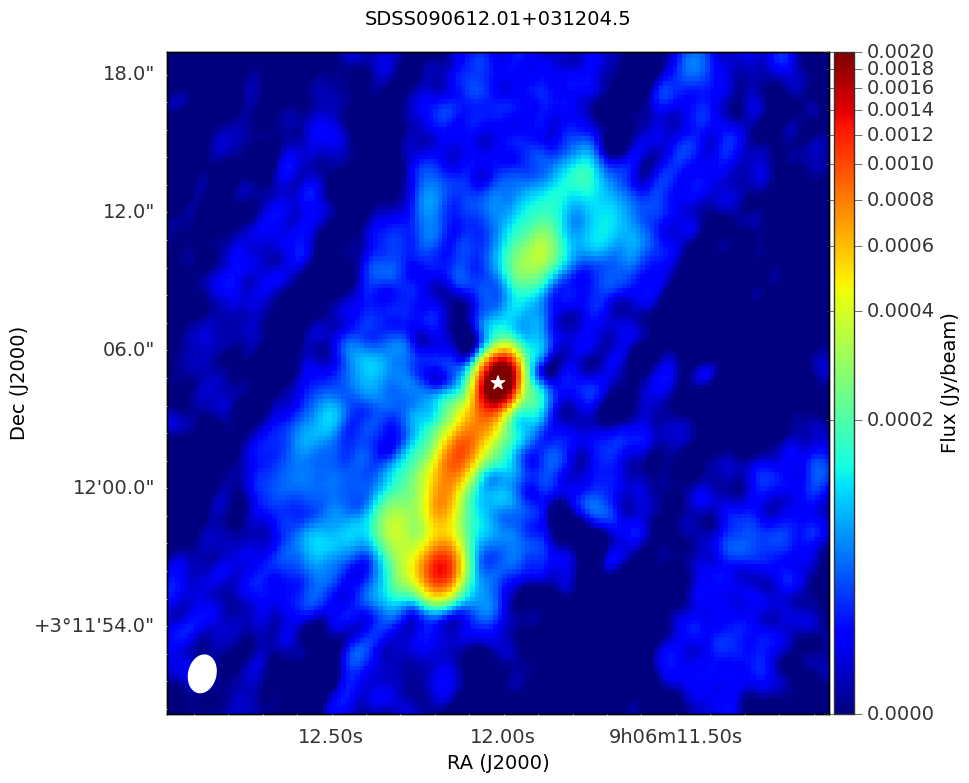}%
\includegraphics[scale=0.17]{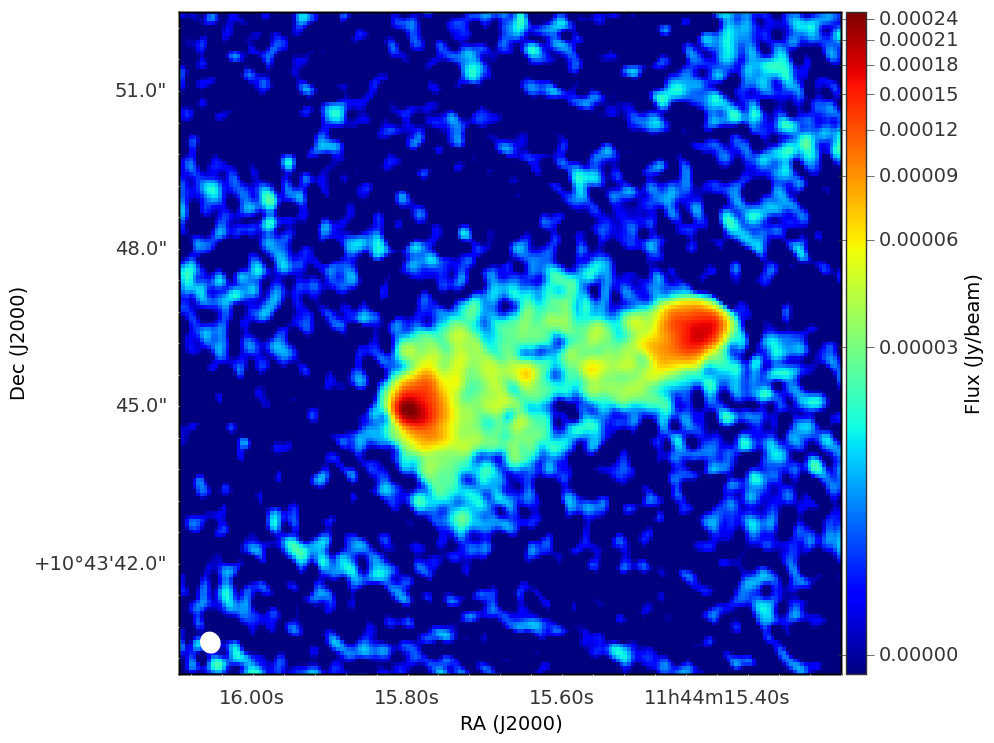}\\
\caption{{\bf Left: }JVLA image of an AGN, SDSS~J090612.01+031204.5.  The beautiful core+lobe structure has a spectral index of $\alpha = -0.76$. It is detected in SDSS with no photo-z measurement and in WISE with a W1$-$W2 colour of 0.83 which confirms its designation as an AGN.  {\bf Right: } JVLA image of a radio lobe source which is undetected in SDSS but detected in WISE.}
\label{pic:beautiful}
\end{figure}

We show two particularly pretty objects from our radio fields in Figure \ref{pic:beautiful}. The source on the left of Figure \ref{pic:beautiful} is a beautiful example of a scaled-down radio core/lobe structure with a clear jet. This source has an SDSS photometric counterpart but no photo-z which prevents an optical classification. There is a WISE detection $<1''$ from the SDSS optical position with W1$-$W2=0.83 mag implying it is an AGN and so it is classified as an ``IR-bright" source in our catalog.  We measure a total 6.0 GHz flux of 10.4 mJy and it is detected at 1.4 GHz in FIRST at 31.52 mJy. While we cannot determine the spectral index of various components as it appears as only a single extended source in FIRST the total spectral index of the entire extended source is $\alpha = -0.76$, suggested the source flux is dominated by the extended lobes.  

The source on the right of Figure \ref{pic:beautiful} has no close optical counterpart in the SDSS catalog though there does appear to be a faint source at the radio location in the SDSS images.  The source leftmost lobe has a radio flux of 1.10 mJy and in the rightmost lobe of 1.24 mJy.  In FIRST at 1.4 GHz it appears that only the right lobe is strongly detected with an integrated flux of 11.06 mJy with the leftmost lobe being visible at a peak flux of 0.69 mJy, a 4.6$\sigma$ detection. This would imply a steep spectral index of $\alpha = -1.5$ in the rightmost lobe, typical of jet-driven radio hotspots and an unusual flat spectral index of $\alpha = 0.32$ in the left lobe. This source is detected in WISE near the position of the radio center and shows W1$-$W2=0.27 mag.  It would thus be placed in the ``IR-bright" category in our field analysis. 

\section{Discussion and conclusions}
\label{sec:conclusion}

In this paper we present a sensitive follow up with JVLA of five low-redshift ($z\sim 0.5$) and eleven high-redshift ($z=2-3$) obscured radio-quiet quasar candidates. 

{\bf Radio properties of the high-redshift type 2 population:} Our high-redshift candidates are selected on the basis of their emission line properties in the SDSS spectroscopy \citep{Alexandroff2013} and in their rest-frame spectra they show moderate levels of obscuration ($\la$ a few mag, \citealt{Greene2014}). The eleven high-redshift quasar candidates that we observed with JVLA are radio-quiet, with typical luminosities $\nu L_{\nu}$[1.4 GHz]$=9.1\times 10^{39}$ erg s$^{-1}$. At redshifts $2-3$, individual radio detections of optically-selected type 2 quasars remain elusive, these objects are radio-quiet and would not be picked up by selection methods which require radio detections to distinguish these objects from dusty star-forming galaxies. Such methods would likely identify only a small fraction of the obscured quasar population. 

The radio and emission line properties of type 2 quasar candidates and extremely red quasar candidates lie on opposite ends of the correlation between radio luminosities and velocity widths of forbidden emission line [OIII]$\lambda$5007\AA\ that exists in $z<1$ AGN and quasars \citep{Mullaney2013, Zakamska2014}. Type 2 quasar candidates are radio-faint and have modest emission line widths (median FWHM[OIII]$=530$ km s$^{-1}$), whereas extremely red quasar candidates which show signatures of fast outflows in their [OIII] emission (FWHM[OIII]$=2930$ km s$^{-1}$) are also significantly brighter in the radio ($\nu L_{\nu}$[1.4 GHz]$=10^{41}$ erg s$^{-1}$). Our observations reinforce the strong relationship between radio emission and kinematics of ionized gas, probing it at high redshifts and wide range of velocity widths. 

{\bf Radio properties of the low-redshift type 2 population.} The mechanics of the radio / gas kinematics relationship and the origin of radio emission in radio-quiet quasars can be difficult to untangle \citep{Mullaney2013, Zakamska2014}. The relationship could be due to either compact jets driving outflows of ionized gas or to radiatively driven winds which produce radio emission as they shock the interstellar medium of the host galaxy. We attempt to address this issue using high-resolution A-array observations of a small exploratory sample of 5 $z<0.8$ radio-quiet and radio-intermediate quasars (four type 2s and one type 1).  Few (if any) radio-quiet sources are spatially resolved when the resolution is improved from a physical scale of $\sim 30$kpc to $\sim 10$ kpc\citep{Hodge2011}, whereas 4 out of 5 are resolved when improving to our resolution of $\sim 2.5$kpc. Thus, galaxy scales might be the key scale to probe in understanding the origin of radio emission in radio-quiet quasars. Future observations of a larger sample of redshift $z < 0.8$ type 2 quasars using the JVLA A-array would confirm this hypothesis. 

Of the five sources observed, one is most likely powered by a jet, and in the JVLA images we are seeing the self-absorbed, compact jet core. Four more objects are spatially resolved and have steep radio spectra in what would be otherwise considered their radio ``core".  In a jet scenario, a steep spectrum core in combination with extended emission requires multiple radio jet cycles every $\sim 10^7$ yr of strong interaction between the jet and the surrounding interstellar medium. Alternatively, we could be seeing radio emission associated with radiatively-driven quasar winds. 
 
{\bf Study of the faint radio population.} We have observed 16 fields with the A-array and the B-array of JVLA in the C-band. Eleven of these fields are also observed in the L-band. Because of the high sensitivity of our observations (typical field center rms of 15 \ujy\ and 80 \ujy\ in the C-band and L-band, correspondingly), in this paper we take an opportunity to investigate the properties of the 179 faint radio sources detected in these observations. 

The nature of the sub-mJy population remains poorly understood. We find SDSS counterparts to 41\% of our radio sources and an additional 33\% are only matched in WISE. Using spectroscopic and photometric diagnostics we classify the counterparts into optical AGN, passive galaxies (with no signs of AGN activity other than the radio source), star-forming galaxies and IR-bright sources (most likely obscured AGN). Although the four classes of counterparts have roughly equal numbers of sources, we suspect that most of star-forming galaxies are mis-identified AGN because the required star formation rate would be implausibly high to account for the observed radio emission. Thus although population models \citep{Massardi2010} predict that the 6 GHz sky below 0.1 mJy should be dominated by star-forming galaxies, we find only a small number of such sources. The fraction of passive galaxies (most likely identifiable with the low-Eddington sources of \citealt{Best2012}) and IR-bright sources is underappreciated in studies of the faint radio sky. 

In the future, the Very Large Array Sky Survey (VLASS\footnote{https://science.nrao.edu/science/surveys/vlass}) will likely share many properties with our study of the sub-mJy radio population.  Current plans for the VLASS will use the B-array at 3.0 GHz down to a co-added sensitivity of 69 \ujy.  Similarly, while sensitive data will be available in some regions, large-scale optical and infrared identification will be accomplished with SDSS and WISE with supplemental data from e.g. PANStarrs, the Dark Energy Survey and Hyper Suprime-Cam. Thus, it is clear from our analysis that it will be challenging to identify optical and infrared counterparts for many of the new faint sources we can expect to detect with the VLASS.  In addition, care must be taken to classify sources based on all available multiwavelength data so as to correctly identify all AGN. While in-band spectral indices will be possible with the VLASS, spectral indices with FIRST will be necessary to probe a large frequency range. In this paper we have raised concerns regarding the statistical analysis of spectral indices for populations in which the flux limit at one frequency is different from the flux limit at the second frequency or in which the sensitivity strongly varies across the mosaics.  Accurately measuring indices in large surveys would require uniform coverage and careful examination of possible biases. Much remains to be discovered about the sub-mJy radio population and a large all-sky survey, especially one that allows for the calculation of spectral indices, will allow us to start definitively identifying the composition of sources contributing in the radio-quiet regime.

\acknowledgements

\clearpage
\LongTables
\begin{landscape}
\begin{deluxetable}{|l|l|l|r|r|r|r|r|r|r|r|}
\tablecolumns{13}
\tablecaption{Full Sample Properties\label{tab:sample}}
\tablehead{\colhead{Field\tablenotemark{1}}&\colhead{RA\tablenotemark{2}}&\colhead{Dec\tablenotemark{3}}&\colhead{F$_{P,6 GHz}$\tablenotemark{4}}&\colhead{rms$_{6 GHz}$\tablenotemark{5}}&\colhead{F$_{P,1.5 GHz}$\tablenotemark{6}}&\colhead{rms$_{1.5 GHz}$\tablenotemark{7}}&\colhead{$\alpha$\tablenotemark{8}}&\colhead{i\tablenotemark{9}}&\colhead{W1\tablenotemark{10}}&\colhead{FIRST\tablenotemark{11}}}
\startdata
SDSS~J0046+0005&11.5355556&0.093513872&1.26E$-$02&3.75E$-$05&6.26E$-$02&2.80E$-$04&$-$0.55&20.013&16.019&86.48\\
SDSS~J0046+0005&11.55222223&0.001347142&3.72E$-$03&0.000241035&2.51E$-$02&5.34E$-$04&$-$0.24&&12.977&29.12\\
SDSS~J0046+0005&11.54333355&0.182291674&5.74E$-$04&7.12E$-$05&1.66E$-$03&5.80E$-$04&$-$0.11&&15.689&1.41\\
SDSS~J0046+0005&11.54244446&0.046513858&2.77E$-$04&2.35E$-$05&1.09E$-$03&3.00E$-$04&$-$0.64&&17.532&1.07\\
SDSS~J0046+0005&11.58750023&0.123513783&2.40E$-$04&3.18E$-$05&7.51E$-$05&1.81E$-$04&1.08&&17.379&\\
SDSS~J0046+0005&11.54933344&0.116291634&1.53E$-$04&2.49E$-$05&$-$1.65E$-$04&1.93E$-$04&NaN&21.734&15.65&\\
SDSS~J0047+0040&11.87120845&0.742138905&1.10E$-$02&9.25E$-$05&5.28E$-$02&1.05E$-$04&$-$0.59&&&58.68\\
SDSS~J0047+0040&11.86126346&0.575138833&8.94E$-$03&5.77E$-$05&1.65E$-$02&8.70E$-$05&$-$0.22&18.559&15.05&19.61\\
SDSS~J0047+0040&11.86915272&0.739750014&4.22E$-$03&9.25E$-$05&1.85E$-$02&1.47E$-$04&$-$0.46&&&22.42\\
SDSS~J0047+0040&11.92282331&0.69152749&2.86E$-$03&1.76E$-$05&4.09E$-$03&6.90E$-$05&$-$0.08&16.558&13.553&3.97\\
SDSS~J0047+0040&11.88143112&0.56719437&7.05E$-$04&5.51E$-$05&1.40E$-$03&8.55E$-$05&$-$0.66&&17.441&0.99\\
SDSS~J0047+0040&11.94349202&0.72463834&3.91E$-$04&2.87E$-$05&6.81E$-$04&7.08E$-$05&$-$0.06&20.932&15.901&\\
SDSS~J0047+0040&11.8659859&0.605249982&3.10E$-$04&1.93E$-$05&5.76E$-$04&7.64E$-$05&$-$0.53&&&\\
SDSS~J0047+0040&11.96738285&0.724860144&2.84E$-$04&4.59E$-$05&3.69E$-$04&6.71E$-$05&0.54&&&\\
SDSS~J0047+0040&11.81370419&0.695360787&2.50E$-$04&1.46E$-$05&2.95E$-$04&7.35E$-$05&0.10&20.843&16.288&\\
SDSS~J0047+0040&11.93143477&0.669082944&2.24E$-$04&1.54E$-$05&4.27E$-$04&6.70E$-$05&$-$0.46&19.244&14.992&\\
SDSS~J0047+0040&11.86776374&0.682527776&1.01E$-$04&1.28E$-$05&2.67E$-$04&1.06E$-$04&$-$0.48&19.793&15.901&\\
SDSS~J0047+0040&11.83165013&0.674694293&6.96E$-$05&8.87E$-$06&$-$2.92E$-$04&7.28E$-$05&NaN&&17.257&\\
SDSS~J0047+0040&11.84659586&0.619527713&6.70E$-$05&1.03E$-$05&6.12E$-$04&1.06E$-$04&$-$3.12&&&\\
SDSS~J0133+0019&23.4072921&0.254058211&1.36E$-$03&2.32E$-$05&4.83E$-$03&7.29E$-$04&$-$0.73&&&5.8\\
\enddata
\tablenotetext{1}{Original Target Field for the source}
\tablenotetext{2}{Full SDSS J2000 right ascension coordinates}
\tablenotetext{3}{Full SDSS J2000 declination coordinates}
\tablenotetext{4}{Peak Flux measured at 6 GHz (mJy/beam)}
\tablenotetext{5}{local RMS at 6 GHz (mJy/beam)}
\tablenotetext{6}{Peak Flux at 1.5 GHz (forced measurement at the location of the brightest component) if available (mJy/beam)}
\tablenotetext{7}{local RMS at 1.5 GHz (forced measurement at the location of the brightest component) if available (mJy/beam)}
\tablenotetext{8}{Spectral Index using tapered 6 GHz peak flux of the brightest component and forced flux measurement at 1.5 GHz}
\tablenotetext{9}{SDSS i$-$band flux if a source match is identified within 1`` (AB mag)}
\tablenotetext{10}{WISE W1 flux (3.6 $\mu$m) if a source match is identified within 3``(Vega mag)}
\tablenotetext{11}{FIRST peak flux at 1.4 GHz if a source match is identified (mJy/beam).}
\end{deluxetable}
\clearpage
\end{landscape}
\bibliographystyle{apj}
\bibliography{quasars_radio}

\end{document}